\documentclass[bibyear]{aa}  
\usepackage{graphicx}
\usepackage{color}

\usepackage[colorlinks=true, linkcolor = blue, urlcolor = magenta, citecolor = blue]{hyperref}

\usepackage{txfonts}

\usepackage{tabularx}

\usepackage{natbib}
\bibpunct{(}{)}{;}{a}{}{,} 
\usepackage{bm}
\usepackage[normalem]{ulem}

\usepackage{placeins}
\usepackage{float}
\usepackage{multicol}

\newcommand{\mat}[1]{\bm{\mathrm{#1}}}

\begin{document}

   \title{{UNIONS}: The impact of systematic errors on weak-lensing peak counts}
   
   \author{Emma Ayçoberry
          \inst{1, 2, 3}
          \and
          Virginia Ajani\inst{2, 4}
          \and
          Axel Guinot\inst{5}
          \and
          Martin Kilbinger\inst{2}
          \and 
          Valeria Pettorino\inst{2}
          \and
          Samuel Farrens\inst{2}
          \and
          Jean-Luc Starck\inst{2}
          \and
          Rapha\"el Gavazzi\inst{3,6}
          \and
          Michael J.~Hudson\inst{7,8,9}
          }

   \institute{Observatoire de Paris, Universit\'e PSL, 61 avenue de l'Observatoire, F-75014 Paris, France
        \and Universit\'e Paris-Saclay, Universit\'e Paris Cit\'e, CEA, CNRS, Astrophysique, Instrumentation et Mod\'elisation Paris-Saclay, 91191 Gif-sur-Yvette, France
        \and
            Institut d'Astrophysique de Paris, UMR 7095, CNRS \& Sorbonne Universit\'e, 98 bis boulevard Arago, F-75014 Paris, France
        \and
            Institute for Particle Physics and Astrophysics, Department of Physics, ETH Zürich, Wolfgang Pauli Strasse 27, CH- 8093 Z\"urich, Switzerland
        \and
            Universit\'e de Paris, CNRS, Astroparticule et Cosmologie, F-75013 Paris, France
        \and
             Laboratoire d'Astrophysique de Marseille, Aix-Marseille Univ, CNRS, CNES, Marseille, France
        \and
            Department of Physics and Astronomy, University of Waterloo, Waterloo, ON, N2L 3G1, Canada
        \and
            Waterloo Centre for Astrophysics, Waterloo, ON, N2L 3G1, Canada
        \and
            Perimeter Institute for Theoretical Physics, 31 Caroline St. N., Waterloo, ON, N2L 2Y5, Canada
            \medbreak
        \email{emma.aycoberry@iap.fr}
   }

   \date{Received XXX; accepted YYY}

  \abstract
  {The Ultraviolet Near-Infrared Optical Northern Survey (UNIONS) is an ongoing deep photometric multiband survey of the northern sky. As part of UNIONS, the Canada-France Imaging Survey (CFIS) provides $r$-band data, which we use to study weak-lensing peak counts for cosmological inference.}
   {We assess systematic effects for weak-lensing peak counts and their impact on cosmological parameters for the UNIONS survey. In particular, we present results on local calibration, metacalibration shear bias, baryonic feedback, the source galaxy redshift estimate, intrinsic alignment, and cluster member dilution.}
   {For each uncertainty and systematic effect, we describe our mitigation scheme and the impact on cosmological parameter constraints. We obtain constraints on cosmological parameters from Monte Carlo Markov chains using CFIS data and \texttt{MassiveNuS} N-body simulations as a model for peak counts statistics.}
   {Depending on the calibration (local versus global, and the inclusion or not of the residual multiplicative shear bias), the mean matter density parameter, $\Omega_\textrm{m}$, can shift by up to $-0.024$ ($-0.5\sigma$). We also see that including baryonic corrections can shift $\Omega_\textrm{m}$ by $+0.027$ ($+0.5 \sigma$) with respect to the dark-matter-only simulations. Reducing the impact of the intrinsic alignment and cluster member dilution through signal-to-noise cuts leads to larger constraints. Finally, with a mean redshift uncertainty of $\Delta \bar z = 0.03$, we see that the shift in $\Omega_\textrm{m}$ ($+0.001,$ which corresponds to $+0.02 \sigma$) is not significant.}
   {This paper investigates, for the first time with UNIONS weak-lensing data and peak counts, the impact of systematic effects. The value of $\Omega_\textrm{m}$ is the most impacted and can shift by up to $\sim 0.03,$ which corresponds to $0.5\sigma$ depending on the choices for each systematics. We expect constraints to become more reliable with future (larger) data catalogs, for which the current pipeline will provide a starting point. The code used to obtain the results is available on GitHub.  \href{https://github.com/CosmoStat/shear-pipe-peaks.git}{\includegraphics[width = 0.017 \textwidth]{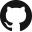}}}
    

    \titlerunning{{UNIONS}: The impact of systematic errors on weak-lensing peak counts}
    \authorrunning{Emma Ayçoberry et al.}

   \maketitle

\section{Introduction}

Weak gravitational lensing has been used as a cosmological probe in recent years with great success, for example with Dark Energy Survey\footnote{\url{https://www.darkenergysurvey.org/}}(DES), Kilo-Degree Survey\footnote{\url{https://kids.strw.leidenuniv.nl/}} (KiDS), Hyper Suprime-Cam\footnote{\url{https://www.naoj.org/Projects/HSC/}} (HSC), and Canada France Hawaii Lensing Survey\footnote{\url{https://www.cfhtlens.org/}} (CFHTLens). It corresponds to the small distortions we observe  in the images of background sources (such as high-redshift galaxies) due to the deflection of photons as they pass through tidal fields in the large-scale structure (LSS) in the Universe \citep{Bartelmann_2001}.

Second-order statistics of weak lensing, such as the two-point correlation function or the power spectrum, only capture the Gaussian part of the LSS \citep{Weinberg-2013}. Its non-Gaussian part, which is induced by the nonlinear evolution of structure on small scales and low redshifts, contains, however, a wealth of information about cosmology. Several higher-order statistics, such as Minkowski functionals \citep{Kratochvil2012,Parroni2020}, higher-order moments \citep[for example,][]{Petri2016, Gatti2020}, the bispectrum \citep{Takada2004,Coulton_2019}, peak counts \citep[and references therein]{Kruse1999,die2010,PhysRevD.91.063507,Lin_2015,Peel_2017,Martinet_2017,li2019,Ajani_2020}, the starlet $\ell_1$ norm \citep{Ajani2021}, the scattering transform \citep{Cheng2020}, wavelet phase harmonic statistics \citep{Allys2020}, and machine learning-based methods \citep[among others]{Fluri2018,Shirasaki2021,2022PhRvD.105h3518F}, have been introduced to account for non-Gaussian information.

In this work we choose to focus on peak counts extracted from real data, using the pipeline developed to perform the study presented in \citet{Ajani_2020}, which had previously only been tested on simulations. Peaks in weak-lensing convergence maps are tracers of overdense regions. They are the local maxima defined as a pixel that is larger than all eight of its neighbors. The peak function -- that is, the number of peaks as a function of peak height (in a convergence map) or signal-to-noise ratio (S/N)-- depends on the nonlinear and non-Gaussian part of the LSS. This higher-order weak-lensing statistic can be used to constrain cosmological parameters. Peak counts are complementary to second-order shear statistics \citep{Jain_2000}, and by combining both parameters degeneracies can be removed \citep{die2010}.

Weak-lensing peaks are an indirect tracer of dark-matter halos: large peaks are strongly correlated with massive halos, whereas lower-amplitude peaks are generally created by multiple smaller halos along the line of sight \citep{yang2011}. Low-amplitude peaks can also be caused by mass outside dark-matter halos, or by galaxy shape noise \citep{Liu2016,yang2011}.

Explicit expressions or complete theoretical predictions of peak counts are still an active area of research. It is, however, possible to generate weak-lensing simulations densely sampled in cosmological parameter space in order to interpolate them and use the interpolation as a prediction. The advantage of simulations is the possibility to incorporate the exact survey mask and shape noise. For example, \citet{die2010} created a set of $N$-body simulations in the $(\Omega_\textrm{m}, \sigma_8)$ plane for $158$ cosmologies, and \citet{liu-2015} and \citet{Kacprzak2016} used ray-tracing $N$-body simulations in the $(\Omega_\textrm{m}, \sigma_8, w )$ plane for $91$ cosmologies. Here we use the \texttt{MassiveNuS} simulations \citep{Liu_2018} to predict peak counts as in \citet{li2019} and \citet{Ajani_2020}. These simulations are described in more detail in Sect.~\ref{sec:simu}.

The first cosmological constraints from peak counts were obtained on real data by \citet{liu-x-2015} using the Canada-France-Hawaii Telescope (CFHT) Stripe 82 Survey, and by \citet{liu-2015}, using CFHTLenS data. DES Science Verification data have been analyzed by \citet{Kacprzak2016}. KiDS 450~deg$^2$ data have been studied by \citet{shan-2017} and \citet{Martinet_2017}. The first tomographic analysis was performed by \citet{harnoisderaps2020cosmic} for the DES-Y1 data release. Recently, \citet{2021arXiv211010135Z} analyzed the DES-Y3 data release using an emulator approach. These analyses all use peak counts and complement second-order statistics analyses, constraining cosmological parameters.

Weak-lensing observables have to be corrected for systematic effects, which can have an observational and astrophysical origin and can induce biases into the cosmological constraints if not properly taken into account. These artifacts can easily be created by the atmosphere, the telescope, and the detector, and during the data analysis. Astrophysical correlations such as intrinsic galaxy alignments add to the lensing correlations in a nontrivial way. Furthermore, to be able to interpret weak-lensing observables in a cosmological context, the physics of small scales needs to be reliably estimated.

Extensive studies of many of these systematics for second-order statistics exist, for example \citet{10.1093/mnras/sts371}, \citet{TROXEL20151}, \citet{Hildebrandt_2016}, and \citet{doi:10.1146/annurev-astro-081817-051928}. For current and future surveys, precise requirements on the instrument and data processing are routinely derived for those statistics. However, how to 
properly include systematic effects in the context of higher-order statistics is an ongoing research topic \citep{Kacprzak2016, Coulton_2020, Z_rcher_2021, Harnois_2021,10.1093/mnras/stab413}.

This paper is the first to study several weak-lensing systematics and uncertainty, their effect on peak counts, and the resulting constraints on cosmological parameters from the Ultraviolet Near-Infrared Optical Northern Survey (UNIONS)/Canada-France Imaging Survey (CFIS) galaxy survey. In addition, we develop a novel method for locally calibrating the measured shear, including shear and selection biases. We investigate mitigation schemes for these effects and quantify biases in cosmological parameters. 

This paper is organized as follows: In Sects.~\ref{sec:data} and \ref{sec:simu} we describe the data and simulations used in this work, respectively. In Sect.~\ref{sec:peaks} we describe the measurement of weak-lensing peak counts. In Sect.~\ref{sec:shear} we introduce local shear calibration and compare it to the standard global calibration. In Sect.~\ref{sec:results} we present results for each of the studied biases and uncertainties and discuss mitigation methods. Finally, we draw our conclusions in Sect.~\ref{conclu}.

\section{Data}
\label{sec:data}

\subsection{Image processing}

We used weak-lensing data from UNIONS\footnote{\url{https://www.skysurvey.cc/}}. This ongoing survey will provide $4,800$~deg$^2$ of multiband photometric images in the Northern Hemisphere. Founded in 2018, UNIONS is a collaboration of several groups and surveys, each providing data in different bands. These participating surveys are CFIS, the Panoramic Survey Telescope And Rapid Response System (Pan-STARRS), Wide Imaging with Subaru HSC of the Euclid Sky (WISHES), and the Waterloo Hawai'i IfA G-band Survey (WHIGS).

For the CFIS part of the survey $r$- and $u$-band images are taken at the CFHT with MegaCAM, a wide-field optical imaging facility. The CFIS $r$-band images have a median seeing of $0.65$ arcsec reflecting the extremely stable atmosphere at Manua Kea together with the excellent CFHT optical system. Each observed sky location is covered by at least three single exposures, which are dithered by one-third of the focal plane, or $0.33^\circ$. The exposure time varies between $100$ and $300$ seconds, where smaller exposure times are chosen in better observing conditions. This survey strategy provides images of a very homogeneous depth.
In this work, we only use $r$-band data from CFIS for our weak-lensing peak count study.
In particular, the weak-lensing data from P3, a patch of size $34.7 \times 17.7$~deg$^2$ is analyzed. This patch overlaps with CFHTLenS, which is used to infer the redshift distribution (see Sect.~\ref{sec:z_estimate}). These data have been processed and validated (see \citealt{UNIONS_Guinot_SP} for details).
The preprocessing of the CFHT data consists of a calibration step with the MegaPipe pipeline \citep{gwyn}. The single exposures are first calibrated before they are combined with SWARP\footnote{\url{https://github.com/astromatic/swarp}} to build stacked images.
MegaPipe provides an astrometric calibration of the survey using \textit{Gaia} Data Release 2 \citep{gaia_dr2}, and photometric calibration of the $r$-band data relative to the Pan-STARRS1 survey  \citep{panstarrs_ps1}. Both astrometric and photometric calibration are excellent so we do not expect calibration errors to significantly impact weak-lensing measurements \citep{UNIONS_Guinot_SP}.

\subsection{The weak-lensing catalog}

Since the stacks have a larger S/N than the single-exposure images, we use the former to detect galaxy candidates. Due to the large dithers, the point-spread function (PSF) on the stacks is very inhomogeneous and discontinuous. For this reason, we detect stars and construct the PSF model on single exposures. Galaxy shapes are then obtained using the multi-epoch model-fitting method \texttt{ngmix} \citep{Sheldon_2017} \footnote{\url{https://github.com/esheldon/ngmix}}. Those measurements are calibrated with \texttt{metacalibration} \citep{huff2017metacalibration} to provide shear estimates. The creation and validation of the shear catalog are fully described in~\citet{UNIONS_Guinot_SP}. We note that this is a preliminary version, called ``version 0,'' of the catalog where the source density is conservative and does not reflect future versions of the CFIS shear data. Specifics of the shear catalog are presented in Table~\ref{tab:survey}.

\begin{table}[t]
\caption{P3-CFIS survey specifications.}
\label{tab:survey}

\begin{tabularx}{\linewidth}{lll}
number density of galaxies & $n_\text{gal}$ & 7 arcmin$^{-2}$ \\ \hline
pixel size & $A_\text{pix}$    &  $0.4^2$ arcmin$^2$/px$^2$ \\ \hline
global ellipticity dispersion & $\sigma_\text{e}$    & $0.44$ \\ \hline
size of the field &   & $34.7 \times 17.7$~deg$^2$ \\
\end{tabularx}

\end{table}

\subsection{Redshift distribution of source galaxies}
\label{sec:z_estimate}

Multiband data of UNIONS are still sparse, and photometric redshifts have not been obtained yet. Using the $r$-band data only,
redshifts have been obtained for a population of galaxies with two different methods, as follows.

For both methods, CFIS galaxies were matched to the deeper CFHTLenS on the $50$~deg$^2$ W3 field.
First, \citet{UNIONS_Guinot_SP} approximated the CFIS redshift distribution as the histogram of the best-fit photometric redshifts from CFHTLenS of that matched subsample. Photo-$z$'s for CFHTLenS had been obtained in \citet{CFHTLenS-photoz} from $u, g, i, r, z$ multiband data, calibrated with various spectroscopic deep data sets.
Second, employing the direct calibration technique, \citet{UNIONS_Spitzer_groups} 
re-weighted the Deep Extragalactic Evolutionary Probe 2 (DEEP2) spectroscopic sample \citep{2013ApJS..208....5N} in a 5D space spanned by the $u, g, i, r, z$ photometric bands, to match the density of the matched subsample in that space. The re-weighted DEEP2 spectroscopic redshift distribution is an estimate of the CFIS $n(z)$. This distribution
was fit in \citet{UNIONS_Spitzer_groups} by an analytical function with two components. The first, exponential, component accounts for the bulk of the distribution,
whereas the second, Gaussian, term, models the tail at $z > 2$. However, the addition of this second term
somewhat overestimates the re-weighted DEEP2 redshift distribution between $z = 1.5$ and $2$.

The mean redshift is obtained as $\bar z = \int_0^{z_\textrm{max}} \textrm{d} z \, z \, n(z)$, where the integral over the normalized redshift distribution $n(z)$ is carried out nominally up to the limiting redshift of the survey, $z_\textrm{max}$. Both redshift distributions have a significantly non-vanishing probability at high redshifts $z \gtrsim 2$. This is most likely not a physical feature, since we do not expect a large number of galaxies in the CFIS sample at those redshifts.

Our best estimate of $\bar z$ is obtained by integrating over the CFHTLenS-matched $n(z)$ with a limit of $z_\textrm{max} = 2$, resulting in $\bar z = 0.65$. Table \ref{tab:zbar} presents alternative estimates with varying $n(z)$ and $z_\textrm{max}$. Three further, reasonable combinations yield a slightly higher $\bar z$ of $0.68$. We use this alternative value in Sect.~\ref{sec:results_z} to test the impact of the estimated redshift uncertainty.

\begin{table}[t]

  \caption{Mean redshift, $\bar z$, of CFIS weak-lensing galaxies for different redshift distributions, $n(z)$, and
maximum redshifts, $z_\textrm{max}$.}
\label{tab:zbar}
\begin{tabular}{llll}
    $n(z)$ & $z_\textrm{max}$ & $\bar z$ & comment 
    \\ \hline
    \cite{UNIONS_Guinot_SP} & $2$ & $0.65$ & fiducial \\
    \cite{UNIONS_Guinot_SP} &  $\infty$ & $0.68$ \\
    \cite{UNIONS_Spitzer_groups} one term & $2$ & $0.68$ \\ 
    \cite{UNIONS_Spitzer_groups} both terms & $2$ & $0.68$ \\ 
    \cite{UNIONS_Spitzer_groups} both terms & $\infty$ & $0.73$ & likely biased \\ 
  \end{tabular}

\end{table}

\section{Simulations}
\label{sec:simu}

To get the predictions for the summary statistics that we use to perform cosmological inference, we employed the \texttt{MassiveNuS} simulations, a suite of cosmological dark-matter-only $N$-body simulations that explore different cosmologies including massive neutrinos in the range $\sum m_\nu = 0$ - $0.62$ eV.
The simulations have a 512 Mpc $h^{-1}$ box size with $1024^3$ cold dark matter particles. The pixel size is $0.4$ arcmin. The implementation is performed using a modified version of the public tree-Particle Mesh (tree-PM) code Gadget2\footnote{\url{https://wwwmpa.mpa-garching.mpg.de/gadget/}} with a neutrino patch, describing the effect of massive neutrinos on the growth of structures up to $k = 10 h$ Mpc$^{-1}$. A complete description of the implementation and the products is provided in  \citet{Liu_2018}. The cosmological parameters vary across the simulations within the range $M_\nu \in [0, 0.62]$,
$\Omega_\text{m} \in [0.18, 0.42]$, and $A_\text{s} \in [1.29, 2.91] \times 10^{-9}$. We thus worked on the constraints on these three cosmological parameters, which are well sampled by the simulations.
They include the effects of radiation on the background expansion and the impact of massive neutrinos is included with a linear-responds method: neutrinos are evolved perturbatively, while their clustering is caused by the nonlinear dark-matter evolution.

The simulations assume a flat universe with Hubble constant $H_0 = 70$ km s$^{-1}$ Mpc$^{-1}$. The primordial power-spectrum scalar index is $n_\textrm{s} = 0.97$, the baryon density $\Omega_\text{b}=0.046$, and the dark-energy equation of state $w=-1$. A fiducial cosmology is set to $[M_\nu, \Omega_\text{m}, 10^{9} \times A_\text{s}] = [0.1, 0.3, 2.1]$. Simulations are available for $101$ cosmologies, with $10000$ realizations for each cosmology, obtained by randomly rotating and shifting the lensing potential planes.

Following \citet{Ajani_2020}, the peak counts were computed for each of the \texttt{MassiveNuS} cosmologies from the simulated convergence maps, averaged over the $10000$ realizations for each model. A model with massless neutrinos corresponding to $[M_\nu, \Omega_\text{m}, 10^{9} \times A_\text{s}] = [0.0, 0.3, 2.1]$ is also provided, and we used it to compute the covariance matrix.

\subsection{Effective redshift distribution}
\label{sec:simulations_z}

For each of the 101 cosmological models, five source redshifts, $z_s=\left\lbrace 0.5, 1.0, 1.5, 2.0, 2.5 \right\rbrace $, are present. To match the simulations to the observed redshift distribution of CFIS, we made the following approximations. We matched the mean redshift between \texttt{MassiveNuS} and CFIS (Sect.~\ref{sec:z_estimate}) and neglected the shape of the redshift distribution. For that, we interpolated the two convergence maps closest in redshift to $\bar z$ and obtained the new effective map as

\begin{equation}
    \kappa_{\bar z} = \kappa_{z=0.5}\lambda + \kappa_{z=1}(1-\lambda).
    \label{eq:kappa_interpolated}
\end{equation}

\noindent This effectively defines the new redshift distribution by a weighted sum of two Dirac delta distributions,

\begin{equation}
    n(z) = \lambda \delta_\textrm{D}(z - 0.5) + (1 - \lambda) \delta_\textrm{D}(z - 1) .
    \label{n_two_delta}
\end{equation}

\noindent To match our best estimate, $\bar z = 0.65$, we set $\lambda = 0.7$. As anticipated in Sect.~\ref{sec:z_estimate}, we also employed $\bar{z}=0.68$ in order to estimate the uncertainty related to the redshift estimation. To do so, we used the relation in Eq.~\ref{eq:kappa_interpolated} to the new effective maps at $\kappa_{\bar z =0.65}$ and $\kappa_{z=1}$ and imposed $\bar{z}=0.68$. This results in an interpolation parameter $\lambda^{\prime} = 0.91$.

\subsection{Noise}\label{subsec:noise}
To include the CFIS shape noise in the simulations that we employ to perform inference, we add Gaussian noise. First, we compute the global ellipticity dispersion $\sigma_\text{e}^2$. The ellipticity noise per smoothed pixel is then

\begin{equation}
    \sigma_{\text{pix}}^2 = \frac{\sigma_\text{e}^2}{2 n_{\text{gal}} A_{\text{pix}}},
    \label{eq:sigma_pix}
\end{equation}where $n_{\text{gal}}$ is the galaxy number density and $A_{\text{pix}}$ is the pixel size. For CFIS data, we used the value listed in Table~\ref{tab:survey}. We applied the Gaussian noise to the convergence maps because we do not have direct access to the simulated shear.

\section{Weak-lensing peak counts}
\label{sec:peaks}
We employed weak-lensing peak counts as summary statistics for our analysis, measured from the weak-lensing convergence maps.

\subsection{Convergence maps}
\begin{figure*}[ht!]
    \centering
    \includegraphics[width=\textwidth]{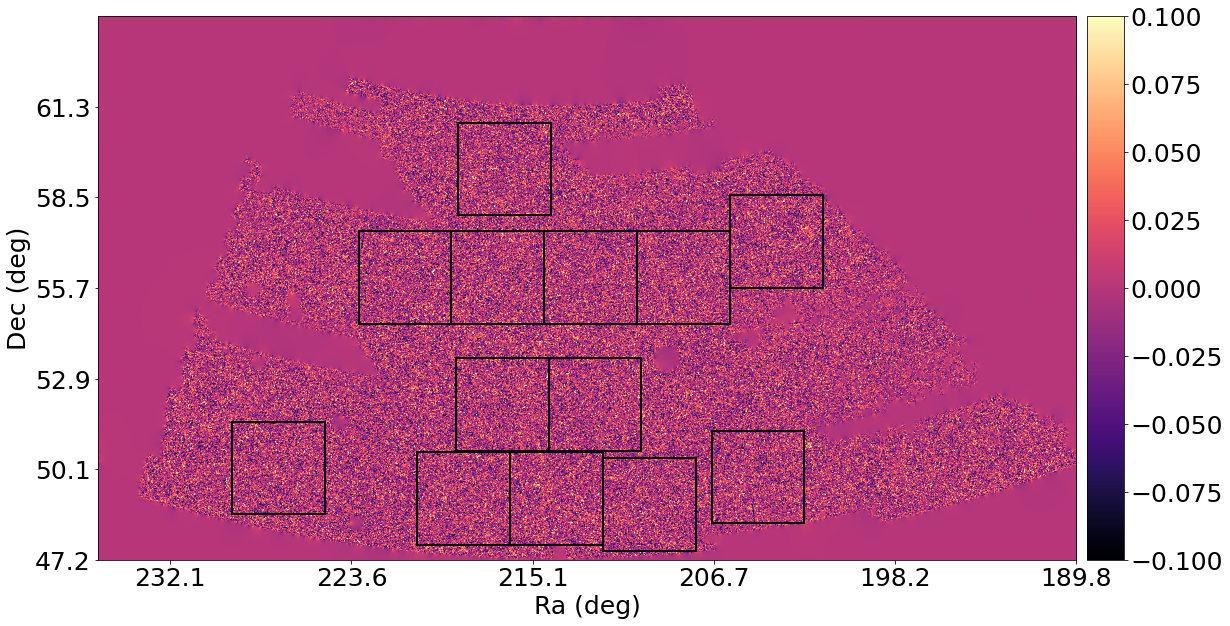}
    \caption{Convergence map of the CFIS-P3 patch. The squares indicate the regions free of large masks, which were used to compute the peak count. The total peak count is the mean of the peaks over the 13 patches.}
    \label{fig:P3_map}
\end{figure*}

The reduced shear $g$ is the main observable in weak gravitational lensing by galaxies, it is estimated from galaxy ellipticities. We can define it as

\begin{equation}
    g = \frac{\gamma}{1-\kappa},
\end{equation}
where $\gamma$ is the shear and $\kappa$ the convergence. For the data, the E- and B-mode convergence maps are built through Kaiser-Squires inversion \citep{ks93} from the reduced shear provided by the weak-lensing catalog. The CFIS-P3 map is quite large for a projection but peaks are local, and any distortion will rather affect large scales. In our analysis we worked with the E-mode of the convergence map as the B-mode contain mostly noise \citep{UNIONS_Guinot_SP}. Hereafter, when we speak about convergence, we mean the E-mode of the convergence map. The CFIS-P3 convergence map is shown in Fig.~\ref{fig:P3_map}. Every pixel has a size of $0.4$ arcmin, as in the simulations. The $13$ black squares are boxes of $512 \times 512$ pixels, which correspond to the size of the simulation convergence maps. These boxes are placed such that they do not overlap with larger masked or missing areas. 

\begin{figure}[]
    \centering
    \includegraphics[width=\hsize]{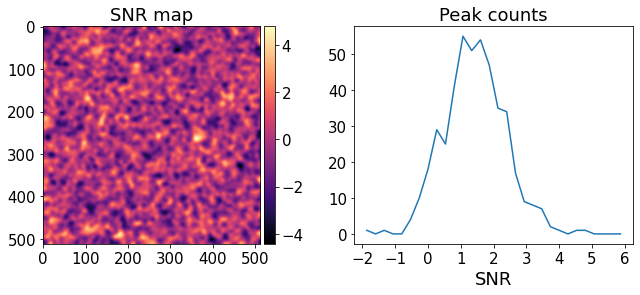}
    \caption{S/N map and peak counts histogram. The left plot is the S/N map of a part of the $512 \times 512$ pixels of CFIS-UNIONS data. The right plot is the peak counts computed on that S/N map.}
    \label{fig:steps}
\end{figure}

For the simulations, we employed the already existing maps from the MassiveNus\footnote{\url{https://saga.edpsciences.org/article/aa/aa43899-22/document/download/iddoc/MassiveNus}} suite, obtained with the LensTools\footnote{\url{https://github.com/apetri/LensTools/blob/master/docs/source/index.rst}} \citep{2016A&C....17...73P} ray-tracing package. We mimicked the CFIS shape noise by adding the noise introduced in Sect.~\ref{subsec:noise} to the maps coming from the simulations, and we smoothed them with a Gaussian kernel of width $\sigma=2$ arcmin. \citet{li2019} find that for the same simulations, using peak counts, the optimal smoothing to get tighter constraints is 2 arcmin. As a first work, we chose this smoothing scale to be consistent with the simulations. Moreover, \citet{KiDS-450} find that a smoothing scale of 2 arcmins is a good choice for  KiDS data, which have the same number of galaxies per pixel as us. For future work, it can be interesting to adapt simulations to the data to determine the optimal smoothing scale for future releases. We thus computed signal-to-noise ratio (S/N) maps, where the S/N is defined as the noisy convergence map, smoothed with a Gaussian filter over the standard deviation of the noise defined in Eq.~\eqref{eq:sigma_pix}. Then we computed the peak counts with the \texttt{lenspack}\footnote{\url{https://github.com/CosmoStat/lenspack}} python package on the S/N maps collecting the local maxima, namely computing the pixels with higher values with respect to their neighboring pixels. In our analysis, we consider linearly spaced bins in the range  S/N$=[-2, 6]$. The peak counts distribution used for parameter inference (Sect.~\ref{sec:results}) corresponds to the mean of the peak counts from the $13$ patches. This corresponds to an area of $13 \times 12.25$ deg$^2 \approx 160$~deg$^2$.
Figure~\ref{fig:steps} presents the S/N map and the peak counts histogram from one patch of the CFIS-P3 data.

\subsection{Modeling and parameter inference}

We modeled the peak function with the \texttt{MassiveNuS} $N$-body and ray-tracing simulations \citep{Liu_2018}. In these simulations the matter density parameter, $\Omega_\textrm{m}$, the primordial power-spectrum normalization amplitude, $A_\textrm{s}$, and the total mass of neutrinos, $M_\nu = \sum m_\nu$, were varied (see Sect.~\ref{sec:simu} for more details).
We constructed a likelihood function as follows. First, the theoretical model was obtained by numerically computing the peak function for each simulated parameter combination. These functions were then interpolated to arbitrary parameters within the parameter boundaries using a Gaussian process. The error of prediction was always below the CFIS statistical error and on the order of a few per cent.
The covariance matrix was computed from the variations of realizations at the fiducial model and is thus assumed to be parameter-independent. As explained in the previous section, for the data, the peaks were computed as the mean of the peaks on 13 mask-free patches; thus, the re-scaling factor of the covariance when we infer parameters from the data is $1/13$. The likelihood was taken as a multivariate Gaussian as a function of the data vector. As prior on the parameters, we used a flat prior for the three parameters: $M_\nu \in [0.06, 0.62]$, $\Omega_\text{m} \in [0.18, 0.42]$ and $A_\text{s} \in [1.29, 2.91]\times 10^9$. The prior of  $\Omega_\text{m}$ and $A_\text{s}$ are the bounds of the simulation. For $M_\nu$, $0.06$ is the minimum from oscillation experiments \citep{ParticleDataGroup:2016lqr} and $0.62$ is the upper bound of the simulations. We explored the posterior distribution with Monte Carlo Markov chain (MCMC) sampling using the python package \texttt{emcee}. Specifically, we employed 250 walkers initialized in a tiny Gaussian ball of radius $10^{-3}$ around the fiducial cosmology $[M_\nu, \Omega_\textrm{m}, 10^9 \times A_\textrm{s}] = [0.1, 0.3, 2.1]$ and estimated the posteriors using 6500 sampling steps and 200 burn-in steps.

\section{Local shear calibration}
\label{sec:shear}

In this section, we describe and measure a spatially varying calibration of the estimated shear. For the multiplicative shear bias, including both shear and selection biases, we use the technique of ``metacalibration'' \citep{huff2017metacalibration}, which we briefly describe in the following subsection.

\subsection{Metacalibration}
\label{sec:metacal}

This method consists in measuring the response matrix $\mat R$ of a shape measurement algorithm to a shear artificially applied to an image. The $i^\textrm{th}$ component of the observed ellipticity of a galaxy, $\varepsilon_i$, is an estimator of the galaxy shear, $\langle \varepsilon_i \rangle = g_i^\textrm{obs}$, however, not an unbiased estimator in general.
The linearized relation between true and estimated shear for a given galaxy can be written as

\begin{equation}
    g_i^{\text{obs}} = \sum_{j=1}^2 R_{ij} g_j^{\text{true}} + c_i,
    \label{eq:g_obs}
\end{equation}

\noindent where the response matrix can be described as $R_{ij} = \partial g_i^{\text{obs}}/\partial g_i^{\text{true}}$ and $c_i$ is the additive bias. The trace of the shear response matrix is also parameterized as $\textrm{tr} \, (\mat R) = 2 (1 + m)$ where $m$ is the multiplicative shear bias.

We numerically compute the shear response matrix by replacing the derivative with finite differences. For this, we create four new images for each detected galaxy. These images are deconvolved with the model PSF at that position, then sheared in two directions for both shear components, and re-convolved with a circularized PSF that is slightly larger than the original one. We note that a fifth image is created, which is unsheared but has the same re-convolved PSF. This
image is used to measure shapes that are consistent with the metacalibration correction.

The response matrix has two additive components, the shear response matrix and the selection response matrix, both of which can be computed via metacalibration, as follows.
First, the shear response matrix is computed for each galaxy individually. Denoting with $g_i^{\textrm{obs}, \pm, j}$ the observed shear of the galaxy image sheared by $\pm \Delta g$ in the $j$ direction, we get

\begin{equation}
    R_{ij}^{\textrm{shear}}
    =
    \frac{g_i^{\text{obs}, +, j} - g_i^{\text{obs}, -, j}}{2\Delta g}
    .
    \label{eq:R_shear}
\end{equation}We used $\Delta g = 0.01$.
Since this is a very noisy measurement, often resulting in singular matrices, we compute the mean value over all galaxies,

\begin{equation}
    \left\langle R^\textrm{shear}_{ij} \right\rangle
    =
    \left\langle
    \frac{g_i^{\text{obs}, +, j} - g_i^{\text{obs}, -, j}}{2\Delta g}
    \right\rangle
    .
    \label{eq:R_shear_av}
\end{equation}

Second, the selection response matrix quantifies selection biases originating from correlations between shear and selection-dependent image properties. For example, if a cut is applied on the S/N of the galaxy sample and the S/N varies with shear, the effective sample shear after the cut is modified, and no longer representative of the underlying population shear, which introduces a bias.

The selection matrix cannot be obtained for a single galaxy, but only for a sample of galaxies. For that, we first apply our selection $S$ to the four samples of sheared galaxies. This provides us with a selection mask $S^{\pm, j}$ for each of the four cases.
These four masks are now in turn applied to the fifth sample without added shear. We compute the mean observed shear for each of the four masked samples, denoted as $\langle g_i^{\textrm{obs}} \rangle_{S^{\pm, j}}$. Any difference in these mean values is due to shear-dependent selection criteria, and can be used to define the selection response matrix as

\begin{equation}
    \langle R_{ij}^{\text{selection}} \rangle =
        \frac{\langle g_i^{\text{obs}} \rangle_{S^{+, j}} - \langle g_i^{\text{obs}}     \rangle_{S^{-, j}}}
        {2\Delta g}.
    \label{eq:R_S}
\end{equation}

\noindent The total mean response matrix is

\begin{equation}
\mat R = \langle \mat R^{\text{shear}} \rangle + \langle \mat R^{\text{selection}} \rangle .
   \label{eq:R_tot}
\end{equation}The estimated shears are calibrated by matrix multiplication with $\mat R^{-1}$, to obtain $g^{\textrm{true}}_i$ for all $i$ following Eq.~\eqref{eq:g_obs}.

In the following sections of the paper, we will need the use the diagonal and off-diagonal terms of the $\mat R^{\text{shear}}$ and $\mat R^{\text{selection}}$ matrix. They are defined as
\begin{eqnarray}
    R_\text{diag} &=& (R_{11} + R_{22})/2 \label{eq:R_diag} \\
    R_\text{off-diag} &=& (R_{12} + R_{21})/2 \label{eq:R_off-diag}
.\end{eqnarray}

\subsection{Global values of the multiplicative shear bias and additive  bias}

Usually, the metacalibration method is applied to the entire catalog, providing us with global values for the shear calibration quantities. We reproduce those results for CFIS-P3 here. The response matrices for CFIS-P3 are

\begin{align}
\mat R^\text{shear} = & \begin{pmatrix} 
\phantom{-}0.74397 & 0.00001 \\ 
-0.00049 & 0.74122 \end{pmatrix} \pm \begin{pmatrix} 
0.00020 & 0.00014 \\ 
0.00022 & 0.00024 \end{pmatrix} ;
\nonumber \\
\mat R^\text{selection} = & \begin{pmatrix} 
-0.10872 & -0.00105 \\ 
-0.00001 & -0.11154 \end{pmatrix}  \pm \begin{pmatrix} 
0.00050 & 0.00046 \\ 
0.00046 & 0.00048 \end{pmatrix} .
\end{align}

\noindent The additive shear bias components are

\begin{equation}
\mat c = \begin{pmatrix} 
-0.00104  \\ 
\phantom{-}0.00043  \end{pmatrix}  \pm \begin{pmatrix} 
0.00011 \\ 
0.00011  \end{pmatrix}
\label{eq:c}
.\end{equation}The errors are computed via jackknife resampling of all galaxies.

\subsection{Local calibration}
\label{sec:local_calibration}

The multiplicative shear bias depends on quantities that vary spatially, such as the PSF properties \citep{2008A&A...484...67P}, galaxy size \citep[for example][]{Spindler-2018, Kuchner-2017} and magnitude \citep{CFHTLenS-shapes}, or the local galaxy density \citep{Henk2017}. These spatial variations may be correlated with shear: in some cases, they both vary with the LSS environment, such as galaxy density and shear. In other cases, residual errors may create cross-correlations, for example, an imperfect PSF calibration influences both calibration and shear.

The understanding and mitigation of spatial patterns in shear calibration is an active field of research for future weak-lensing surveys \citep{Kitching_2021,2022arXiv220301460C}.
Here, we investigate local shear calibration, the dependence of calibration on observed quantities, and the impact on cosmological parameters.

In the context of weak-lensing mass maps as tracers of the LSS, \citet{Van_Waerbeke_2013} carry out a local calibration in pixels of $1$ arcmin size.  Their multiplicative shear bias is computed as a smooth, two-parameter fitting function from image simulations. 
This is in contrast to our case of metacalibration, which provides very noisy calibration estimates for individual galaxies. These estimates need to be averaged over substantially larger areas to reduce the uncertainty such that the corresponding response matrices are numerically stable enough for inversion. 

In this work, we employed a series of square patches with a size of $d_{\textrm{s}} = 4, 2, 1, 0.5$, and $0.25$ degrees over which the shear calibration estimates are averaged in turn. To easily divide the observed sky area into an integer number of square-sized sub-patches, we first projected the data into a Cartesian plane. Next, we extended this area via zero padding to size $N_x \times N_y$~deg$^2$, such that $N_x$ and $N_y$ are multiples of the largest sub-patch size $d_\textrm{s, max} = 4$ deg.

We computed local versions of $\mat R^\textrm{shear}$ \eqref{eq:R_shear_av} and
$\mat R^\textrm{selection}$ \eqref{eq:R_S} by carrying out the averages per sub-patch. If the number of galaxies in a sub-patch was smaller than a threshold $n_\textrm{gal, 0}$, we replaced the local response matrices in that sub-patch by their global mean, to avoid numerical instabilities.
We chose $n_\textrm{gal, 0} = \bar n_\textrm{gal} / 2$. This occurs mainly at the edges of the field where sub-patches overlap with the area of zero padding.
We calibrated the estimated shear of each galaxy in a given sub-patch by the total local response matrix, in analogy to \eqref{eq:R_tot}. Similarly, we computed the local shear bias $\vec c$ as the average per sub-patch.

We computed the error of the local response matrices by creating jackknife resamples from the smallest sub-patch size $d_\textrm{s, min} = 0.25$ deg. Each sub-patch of size $d_\textrm{s} > d_\textrm{s, min}$ was thus split into $(d_\textrm{s} / d_\textrm{s, min})^2$ jackknife subsamples; for the smallest sub-patch size, we could not compute the error. The additive bias is easily obtained locally by computing jackknife errors over galaxies per sub-patch.

\subsection{Results of the local calibration}

\subsubsection{Variation of the shear and selection response}

\begin{figure*}[ht]
    \centering
    \includegraphics[width=\textwidth]{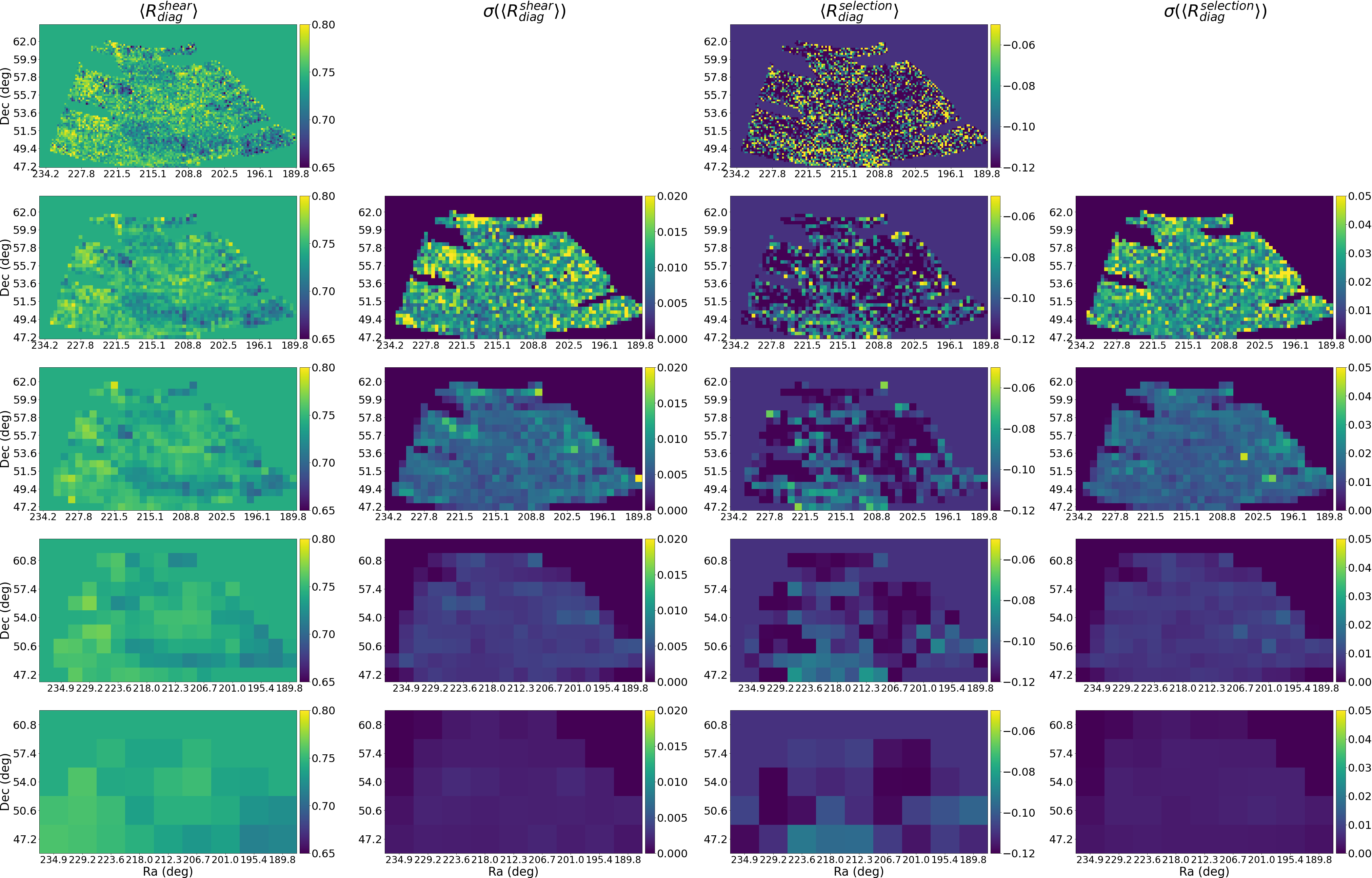}
    \caption{$\langle R_\text{diag} \rangle$ depending on the size of the local calibration. From top to bottom, calibration is done on $0.25, 0.5, 1, 2, 4$~deg$^2$. The four columns are $\langle R_\text{diag}^\text{shear} \rangle$, its standard deviation, $\langle R_\text{diag}^\text{selection} \rangle$, and its standard deviation.}
    \label{fig:local_calib_plot}
\end{figure*}

We show the spatial variation of $\mat R^{\text{shear}}$ and $\mat R^{\text{selection}}$ in sub-patches of different size in Fig.~\ref{fig:local_calib_plot}.
From top to bottom the shear and selection responses are computed on sub-patches with increasing size of $d_\textrm{s} = 0.25, 0.5, 1, 2$ and $4$ degrees. The smaller the sub-patch size, the more the response shows variations corresponding to the local spatial environment. The larger the sub-patch size, the more the calibration approximates the global calibration.
The errors of the shear response are small, even in the case of the $0.5$ degree sub-patch size. Concerning the selection matrix, we can see that on $0.5$~deg$^2$, the errors have the same order of magnitude as $\mat R^\textrm{selection}$. On $1$~deg$^2$, the errors are smaller than $\mat R^\textrm{selection}$ indicating that the computation of the local selection calibration is precise when done on $1$~deg$^2$ or more. To emphasize these conclusions, the 1D distributions of $\mat R^\textrm{shear}$ and $\mat R^\textrm{selection}$ are plotted in Fig.~\ref{fig:hist_R}.

\subsubsection{Variation of the additive bias}

\begin{figure*}
    \centering
    \includegraphics[width=\textwidth]{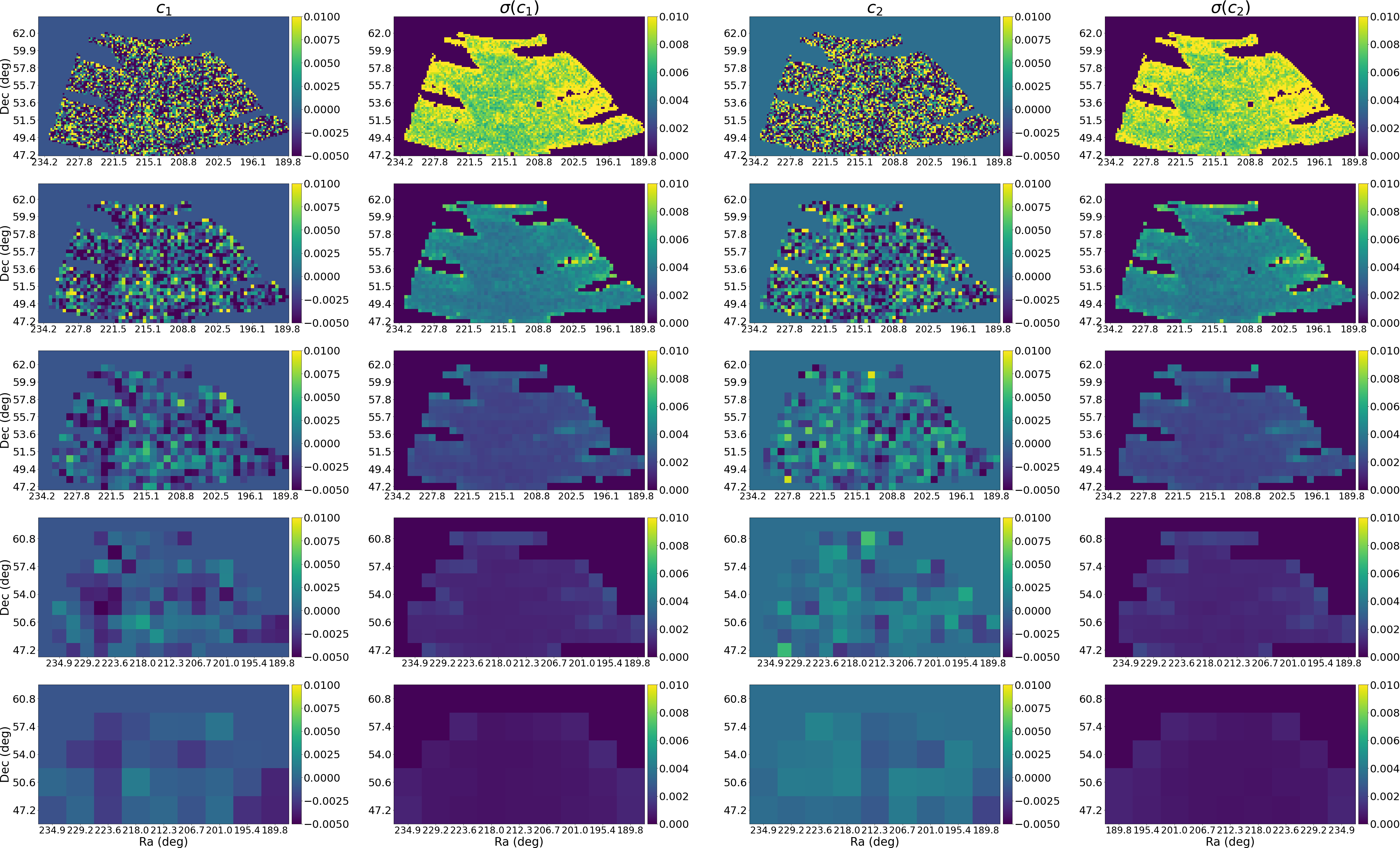}
    \caption{$c_1$ and $c_2$ depending on the size of the local calibration. From top to bottom, calibration is done on $0.25, 0.5, 1, 2, 4$~deg$^2$. The four columns are $c_1$, its standard deviation, $c_2$, and its standard deviation.}
    \label{fig:c}
\end{figure*}

The additive bias and its standard deviation are shown in Fig.~\ref{fig:c}. The standard deviation is as large as the value itself for local calibration on small scales. Except for the largest sub-patch sizes, the values of $c_1$ and $c_2$ per pixel are consistent with zero. This reflects the large fluctuations and randomness of the additive bias on small scales. To emphasize these conclusions, the 1D distributions of $c_1$ and $c_2$ are plotted in Fig.~\ref{fig:hist_c}.

\subsubsection{Relative errors}
\begin{figure*}
    \centering
    \includegraphics[width=\hsize]{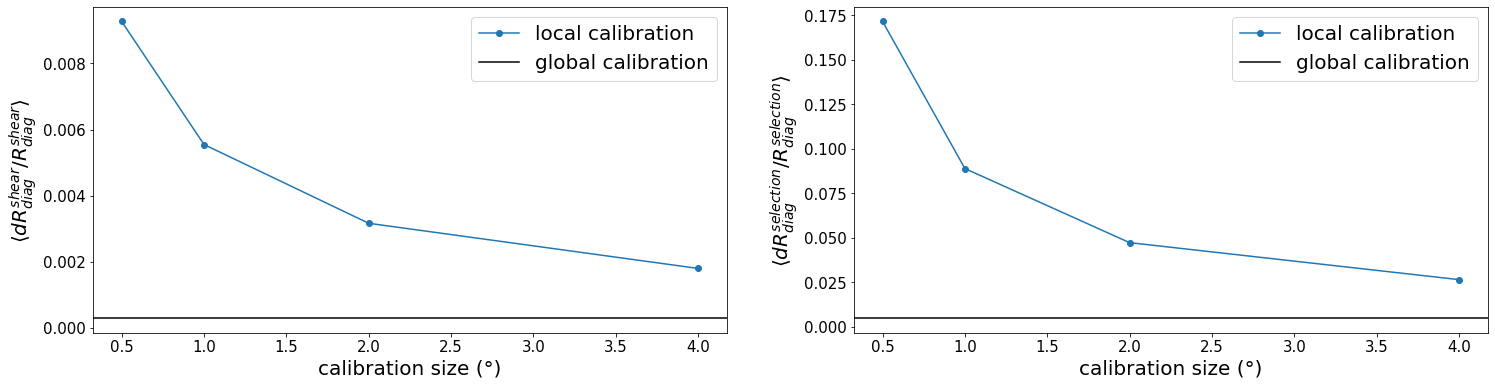}
    \caption{Relative errors of shear matrix (\emph{left panel}) and selection matrix (\emph{right panel}). In both cases, the blue line is the relative error of the diagonal terms of the matrix depending on the calibration size. The black line is the relative error of the diagonal terms of the matrix computed with the value obtained on the global calibration case.}
    
    \label{fig:relative_errors}
\end{figure*}

We computed the errors of the response matrices at different sub-patch sizes, averaged over all sub-patches (blue lines), compared to the relative errors of the global calibration (black lines). The results are shown in Fig.~\ref{fig:relative_errors}.

The relative error of the shear matrix is below 1\% in all cases but always above the relative error of the global calibration, meaning that there are fluctuations at all the studied scales, but these are small. The fractional error of the selection matrix is higher. It reaches $17\%$ at small sub-patch sizes but decreases up to ~3\% at 4~deg$^2$. For the shear and selection matrix, we see that when we go on a smaller calibration size, the relative errors asymptotically approach the one of the global calibration.

\subsection{Parameter correlation matrix} 

\begin{figure*}
    \centering
    \includegraphics[width=\textwidth]{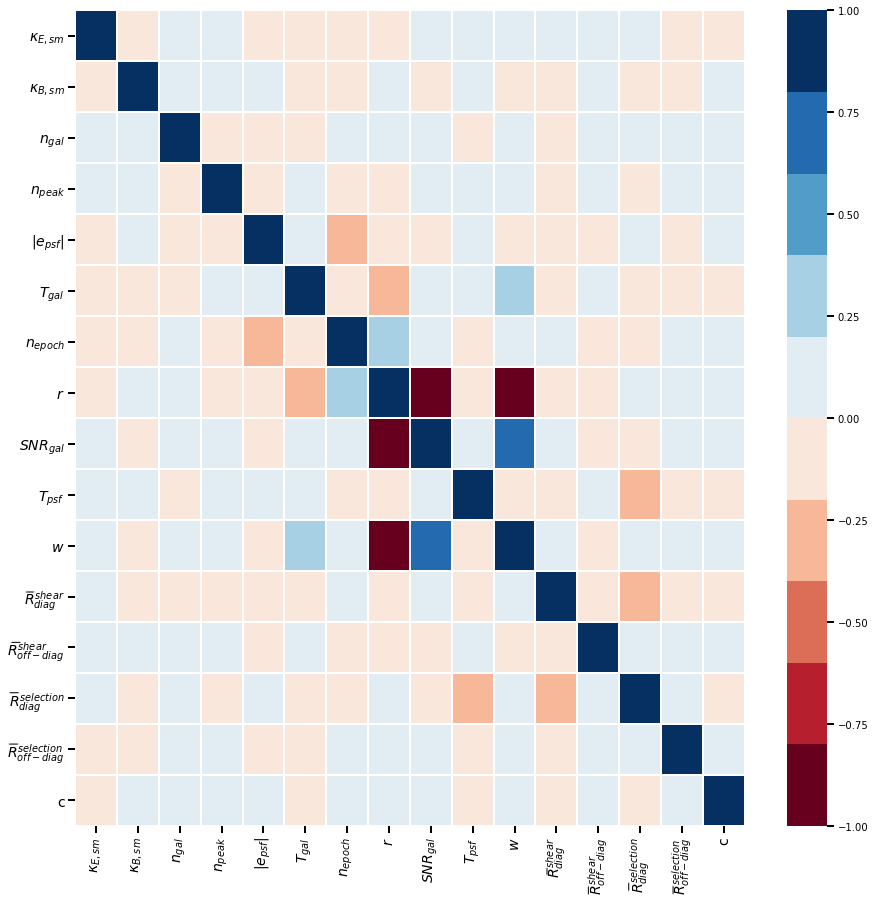}
    \caption{Correlation matrix of the quantities listed in Table \ref{tab:corr}. The colors indicate the amplitude of the correlation, ranging between $-1$ and $1$. All quantities are mean values.}
    \label{fig:corr}
\end{figure*}

The calculation of the local shear bias allows us to further
explore possible origins of shear bias, and their influence on other quantities obtained from the galaxy sample. To that end, we computed the correlation matrix between different quantities, for which we used patches of size $d_\textrm{s} = 1$ deg. We also considered correlations between quantities other than shear bias. The correlation matrix is calculated using the \texttt{pandas} \textit{DataFrame.corr} function\footnote{\url{https://pandas.pydata.org/docs/reference/api/pandas.DataFrame.corr.html}}; it uses the Pearson method, which computes the standard correlation coefficient. The correlation matrix between the different quantities listed in Table \ref{tab:corr} is shown in Fig.~\ref{fig:corr}. To better see the correlation terms by terms, another correlation matrix is shown in Fig.~\ref{fig:corr_app}.

\begin{table}[ht!]

\caption{Symbols and description of quantities used in the cross-correlation matrix (Fig.~\ref{fig:corr}). }

\label{tab:corr}

\begin{tabularx}{\linewidth}{l|X}
Symbol & Description \\ \hline
$\kappa_{E, \textrm{sm}}, \kappa_{B, \textrm{sm}}$     & Smoothed E- and B-mode convergence, respectively \\
$n_\textrm{gal}$     &  number of galaxies per pixel \\
$n_{\textrm{peak}}$ & number of peaks \\
$|e_\textrm{psf}|$  & modulus of the point spread function \\
$T_\textrm{gal}$ & galaxy size \\
$n_{\textrm{epoch}}$ & number of single-exposure epochs \\
$r$ & $r$-band galaxy magnitude \\
SNR$_\textrm{gal}$ & galaxy signal-to-noise ratio \\
$T_\textrm{psf}$ & psf size \\
$w$ & galaxy weight \\
$\overline R_\text{diag}^{\textrm{shear}}$, $\overline R_\text{off-diag}^{\textrm{shear}}$ & Average of the diagonal and off-diagonal shear response matrix element, respectively, Eq.~\eqref{eq:R_shear_av} \\
$\overline R_{diag}^{\textrm{selection}}$, $\overline R_\text{off-diag}^{\textrm{selection}}$ & Average of the diagonal and off-diagonal selection response matrix element, respectively, Eq.~\eqref{eq:R_S} \\
$c$ & Average of the additive bias element\\
\end{tabularx}

\end{table}

First, we see that the E- and B-mode smoothed convergence, $\kappa_{E, \textrm{sm}}$ and $\kappa_{B, \textrm{sm}}$, are not strongly correlated. This indicates that there is no significant systematic effect that mixes both modes. Neither mode is strongly correlated to other quantities. Also, the number of peaks $n_\textrm{peak}$ is not correlated to observational effects.

Further, there is no visible correlation between the PSF size and the additive bias. This is evidence for the correct estimation of PSF size since a bias in the PSF size typically leads to an additive shear bias. We also note the absence of a correlation between the PSF ellipticity and the shear selection matrix.
This gives us confidence that the PSF deconvolution in the metacalibration process works well.
There is, however, a $20$ - $30$\% negative correlation between the selection response elements and the size of the PSF. A plausible explanation is that in areas with larger seeing, fewer small and faint galaxies make it into the weak-lensing sample. This leads to a stronger (more negative) selection bias, which is reflected in the anticorrelation.

A negative correlation can be seen between the number of epochs, $n_\textrm{epoch}$, and the PSF ellipticity modulus $| e_\textrm{PSF}|$. Since the former is the mean over all contributing epochs, a reason for this anticorrelation might be the PSF ellipticity gets more circular when averaged over more independent observations.

There is a negative correlation of the $r$-band magnitude with galaxy weight $w$, S/N, and size $T_\textrm{gal}$, as well as a positive correlation with $n_\textrm{epoch}$. This reflects the expectation that faint galaxies have a lower weight and S/N, and are easier observed with more exposures.

\section{Impact of uncertainty and systematic effects on cosmological parameters}
\label{sec:results}

This section explores different systematic effects and uncertainties, and their impact on cosmological parameter constraints. We quantify the potential biases on cosmology 
from the spatially varying shear calibration (Sect.~\ref{sec:result_local}), the redshift uncertainty (Sect.~\ref{sec:results_z}),
a residual shear bias (Sect.~\ref{sec:delta_m}), baryonic feedback (Sect.~\ref{sec:baryon}), intrinsic galaxy alignment and cluster member dilution (Sect.~\ref{sec:further_effects}. All these effects are studied jointly in Sect.~\ref{best_case}.

For each effect, we will show constraints on cosmological parameters as compared to those obtained with the ``ideal'' case. This case, which is represented in blue in all the following figures represents the constraints obtained when we use the data calibrated globally without residual shear bias, a mean redshift of $\bar z = 0.65$, without baryonic correction, cluster member dilution or intrinsic alignment. This assumes an unrealistic best-case scenario of vanishing spatial variation of shear calibration and residual shear bias, and no biases from baryons, intrinsic alignment or cluster members. When the parameters are well-constrained, we specify the shift compared to the ideal case, which is the case for $\Omega_\textrm{m}$. The parameters $M_\nu$ and the lower bound for $A_\textrm{s}$ are in most cases not well constrained within the prior.

\subsection{Local calibration}
\label{sec:result_local}

\begin{figure}
    \centering
    \includegraphics[width=\hsize]{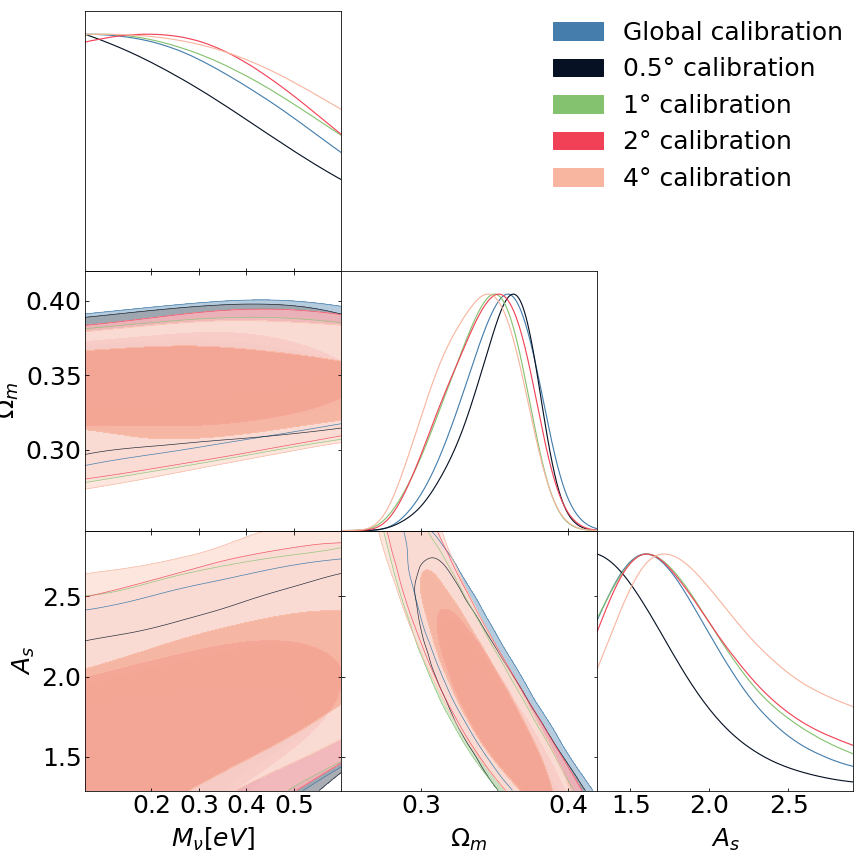}
    \caption{1D and 2D marginal posteriors for CFIS-P3 using different calibration sizes in the metacalibration. The 2D inner and outer contours show the $68\%$ and $95.5\%$ credible region, respectively. The case when the calibration is done globally (blue) is compared to the case where the calibration is done on 0.5 (black), 1 (green), 2 (red), and 4 deg$^2$ (pink).~ $\Omega_\textrm{m}$ shifts by $-0.015,$ which corresponds to $-0.3\sigma$ when the calibration goes from global to local on $4$~deg$^2$.}
    \label{fig:P3_corr_z065_all_deg}
\end{figure}

To study the impact on cosmology of spatially varying shear biases, we performed a local shear calibration using different scales, following the method explained in Sect.~\ref{sec:local_calibration} for the multiplicative shear bias. For the additive bias, we used the value obtained with the global calibration because the local calibration value is very noisy. The result is shown in Fig.~\ref{fig:P3_corr_z065_all_deg}. We see that a local calibration will always shift $\Omega_\textrm{m}$ to a lower value except for the calibration on $0.5$~deg$^2$. These variations are all within the statistical error bars. There is no systematic variation that becomes evident when going to smaller calibration sizes. For $A_\textrm{s}$ and $\sum M_\nu$, the trend is not clear. $\Omega_\textrm{m}$ shifts by $-0.015,$ which corresponds to $-0.3\sigma$ (i.e.,~$0.3$ times the statistical uncertainty) when the calibration goes from global to local on $4$~deg$^2$ for example. 

From this work, it is not clear which calibration size is the best one because, depending on the size of the calibration, we will capture more or less the local effect. Future work has to be done to determine the criteria to choose the size of the calibration. In this work, when we need a local calibration, we use the one done on $1$~deg$^2$ because the relative error of the selection matrix (which is the limiting one) is below 10\%.

\subsection{Residual multiplicative shear bias}
\label{sec:delta_m}

The shear bias $m$ computed earlier with metacalibration is not perfect. A residual bias $\Delta m = m^{\textrm{metacal}} - m^{\textrm{true}}$ remains and generally has to be quantified with image simulations. We used the value $\Delta m = 0.007$ based on the results from \citet{UNIONS_Guinot_SP}, who found $\Delta m = 0.007 \pm 0.03 $, estimated using CFIS-like image simulations of isolated galaxies and ignoring the effect of blending. In comparison, other surveys also find residual biases at around the per cent level or below; for example, \citet{maccrann2020des} and \citet{Gatti_2021} state a value of $\Delta m = -0.0208 \pm 0.0012$ for DES-Y3.

\begin{figure}
    \centering
    \includegraphics[width=\hsize]{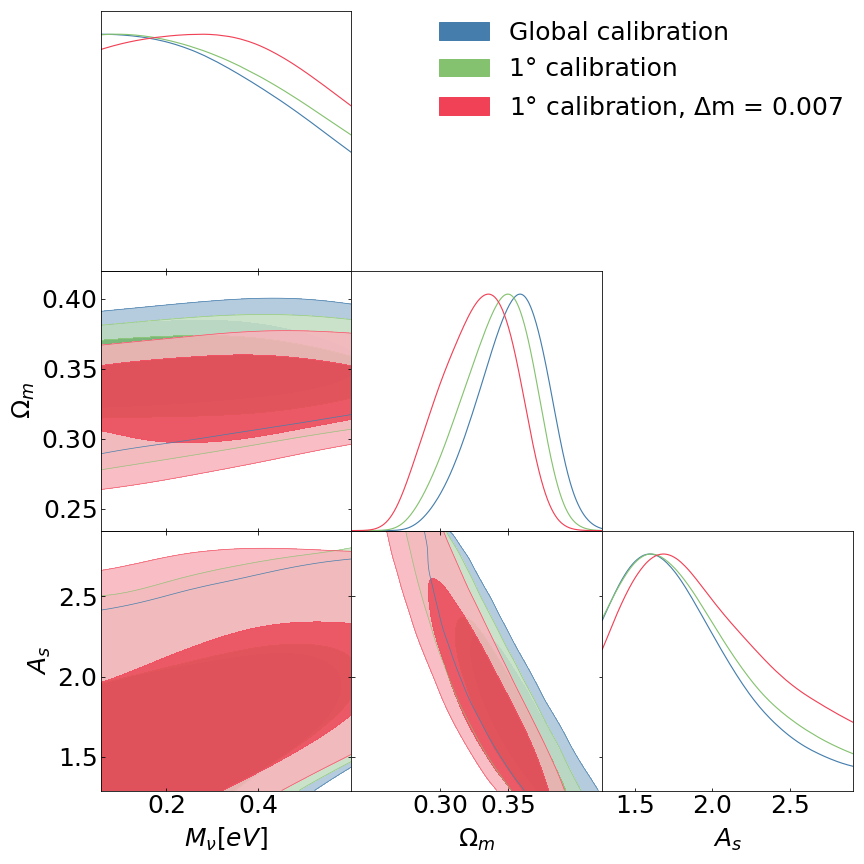}
    \caption{1D and 2D marginal posteriors for CFIS-P3, with the residual multiplicative shear bias added. The 2D inner and outer contours show the $68\%$ and $95.5\%$ credible region, respectively. $\Omega_\textrm{m}$ shifts by $-0.01,$ which corresponds to $-0.2 \sigma$ when the calibration goes from global (blue contour) to local on $1$~deg$^2$ (green contour). When $\Delta m$ is added (red contour), $\Omega_\textrm{m}$ shifts by $-0.024,$ which corresponds to $-0.5 \sigma$ with respect to the global calibration.}
    \label{fig:constraint_1deg}
\end{figure}

We modeled the effect of the residual multiplicative shear bias by adding the residual bias $\Delta m$ to the response matrix in the local case. This corresponds to a conservative, worst-case scenario where the residual bias is constant in space. Thus, Eq.~\eqref{eq:R_tot} is modified to

\begin{equation}
\mat R = \langle \mat R^{\text{shear}} \rangle + \mat I \times \Delta m + \langle \mat R^{\text{selection}} \rangle ,
\end{equation}where $\mat I$ is the identity matrix. The impact of this bias on the cosmological constraints is shown in Fig.~\ref{fig:constraint_1deg}. In blue, the calibration is global whereas it is done on $1$~deg$^2$ for the green and red cases. Moreover, the red case includes the multiplicative shear bias. Including the residual shear bias changes the values of the parameters in the same direction as the local calibration compared to the global calibration. $\Omega_\textrm{m}$ shifts by $-0.024,$ which corresponds to $-0.5 \sigma$ with respect to the global calibration. A positive $\Delta m$ effectively reduces the estimated shear, resulting on average in a smaller number of peaks, and thus smaller clustering parameters. The 1D marginalized contours for $A_\textrm{s}$ shift to the right, contrary to what we would expect. However, the joint $\Omega_\textrm{m}$ - $A_\textrm{s}$ 2D contours clearly reflect the smaller calibrated shear and shift toward smaller clustering amplitudes.

\subsection{Redshift uncertainty}
\label{sec:results_z}
\begin{figure}
    \centering
    \includegraphics[width=\hsize]{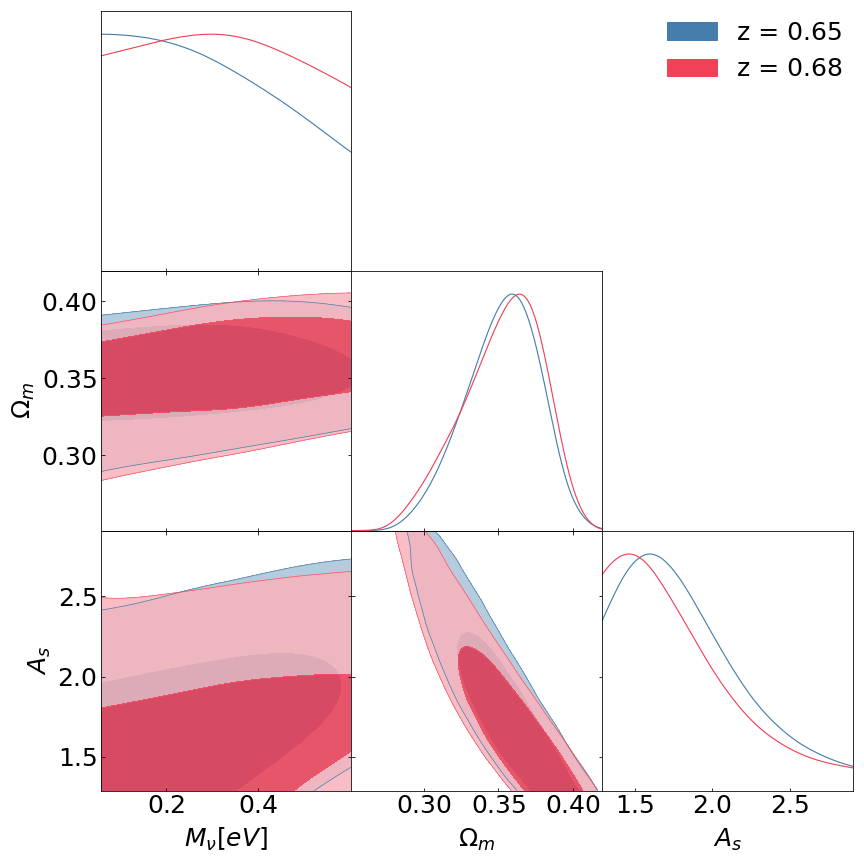}
    \caption{1D and 2D marginal posteriors using simulations at different redshift. The 2D inner and outer contours show the $68\%$ and $95.5\%$ credible region, respectively. The shear calibration is global. $\Omega_\textrm{m}$ shifts by $+0.001,$ which corresponds to $+0.02 \sigma$ when the redshift goes from 0.65 (blue) to 0.68 (red).}
    \label{fig:constraints_delta_z}
\end{figure}

As discussed in Sect.~\ref{sec:z_estimate}, $\bar z = 0.65$ is our best estimate of the redshift of the data observed from CFIS. This estimate has however a level of uncertainty. To quantify the impact of this uncertainty on our results, we compared cosmological constraints using the simulated convergence maps interpolated to $\bar z = 0.65$ and $\bar z = 0.68$, which corresponds to the two mean redshift estimates discussed in Sect.~\ref{sec:z_estimate}. The result is shown in Fig.~\ref{fig:constraints_delta_z}. Fitting the data with the model at a higher mean redshift only slightly shifts the posterior: $\Omega_\textrm{m}$ shifts by $+0.001,$ which corresponds to $+0.02 \sigma$ when the redshift goes from 0.65 to 0.68. These shifts are well within the statistical uncertainties of the two parameters. Such a shift is, however, expected to have a significant influence on analyses using a larger survey area and/or redshift tomography. We also performed a test using only simulations, where the data vector was the mean of the fiducial simulation at $z = 0.65$ or at $z = 0.68$. The result is presented in Appendix \ref{app:sim}. We note, however, that when using simulations only, the shift appears to be larger, this may indicate some residual systematics that impacts the results on the data shown in this section.

\subsection{Baryonic feedback}
\label{sec:baryon}

The impact of baryons on the LSS is important for cosmological analyses with weak-lensing peak counts \citep{Osato_2015,harnoisderaps2020cosmic,Coulton_2020}. The redistribution of matter due to baryonic processes tends to reduce the number of high S/N peaks and augment that of smaller S/N values.

We used the results from \citet{Coulton_2020} who model the fractional difference of the number of peaks, $\Delta N_\textrm{peaks}/N_\textrm{peaks}$, for three different baryon physics scenarios based on the \texttt{BAHAMAS} hydrodynamical simulations \citep{McCarthy-2016} compared to \texttt{MassiveNuS} dark-matter-only simulations \citep{Liu_2018}. This assumes that baryonic processes are independent of the underlying cosmology. The three scenarios are denoted as LowAGN, Fiducial, and HighAGN, where the amount of baryonic feedback increases in that order. The fiducial baryonic correction comes from the \texttt{BAHAMAS} simulations with a feedback model designed to best match the observations. The LowAGN and HighAGN corrections are simulations where the active galactic nucleus (AGN) heating is lowered or raised by 0.2 dex, respectively,  and thus the simulations skirt the lower and upper bounds of the observed gas fraction. 

\begin{figure}
\centering
    \includegraphics[width=\hsize]{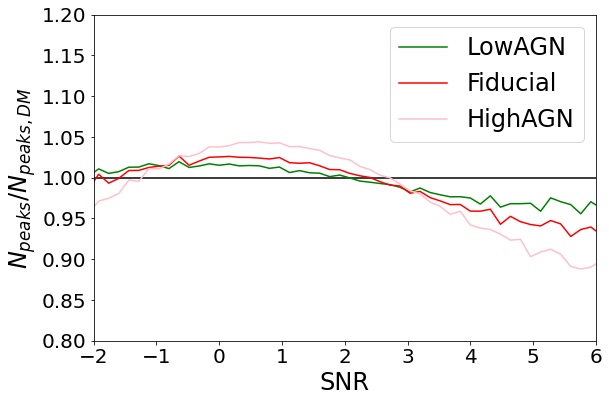}
    \caption{Fractional difference in peak count obtained with \citet{Coulton_2020} data. The difference between simulations with LowAGN, Fiducial, or HighAGN and dark-matter-only simulations is presented in green, red, and pink, respectively.}
    \label{fig:reprod_coulton}
\end{figure}

We modified the model predictions of peak counts by multiplying the number of peaks from the \texttt{MassiveNuS} (dark-matter-only) simulations by $N_{peaks}/N_{peaks,DM}$, shown in Fig.~\ref{fig:reprod_coulton}.

\begin{figure}
    \centering
    \includegraphics[width=\hsize]{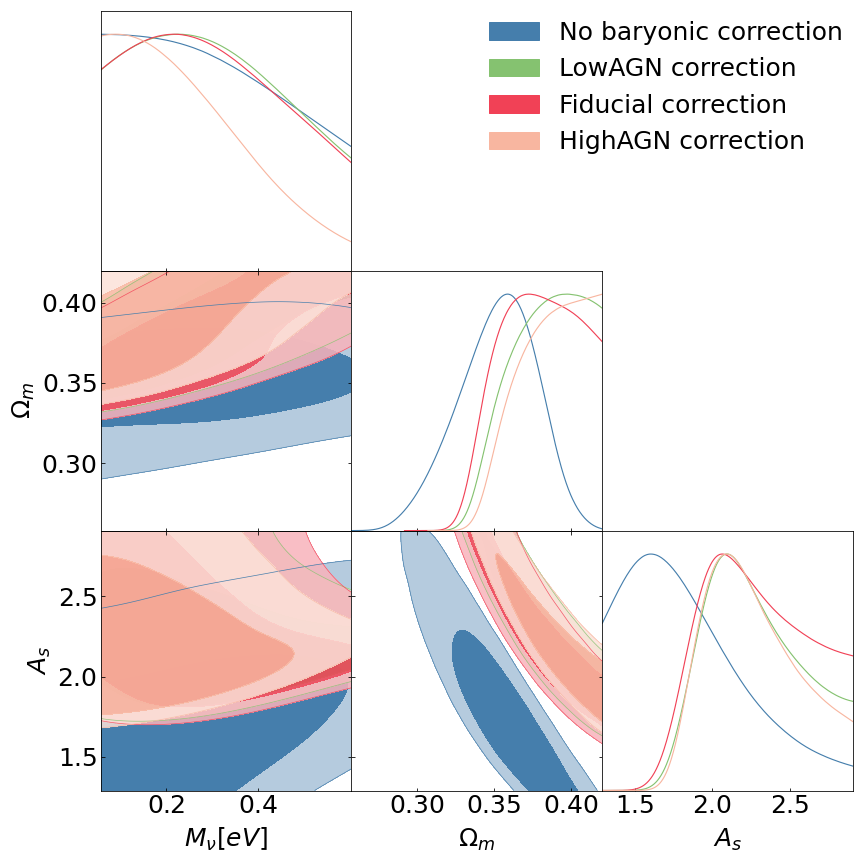}
    \caption{1D and 2D marginal posteriors for CFIS-P3, with the peak count predictions corrected with different baryonic feedback. The 2D inner and outer contours show the $68\%$ and $95.5\%$ credible region, respectively. The case without baryonic correction (blue) is compared to the three baryonic scenarios: LowAGN (green),  Fiducial (red), and HighAGN (pink). $\Omega_\textrm{m}$ shifts by $+0.027,$ which corresponds to $+0.5 \sigma$ when the fiducial goes from dark-matter-only to LowAGN correction.}
    \label{fig:P3_corr_z065}
\end{figure}

We can see how the constraints evolve when we use real CFIS-P3 data and corrected simulations. To see the effect of the baryonic correction only, we use the global calibration, simulations at $z = 0.65$, no intrinsic alignment or cluster member dilution correction. We do not know how the baryonic feedback influences the observed data thus we correct the \texttt{MassiveNuS} simulations with the three different baryonic feedback corrections. In Fig.~\ref{fig:P3_corr_z065} we show the results of these corrections (LowAGN in green, Fiducial in red, and HighAGN in pink) compare to the case of the dark-matter-only simulation (blue).

Baryonic feedback has a significant impact on the best-fit parameters.
$\Omega_\textrm{m}$ and $A_\textrm{s}$ are shifted
to higher values with increasing baryonic modifications. For example, $\Omega_\textrm{m}$ shifts by $+0.027,$ which corresponds to $+0.5 \sigma$ when the fiducial goes from dark-matter-only to LowAGN correction. \citet{li2019} found that it is the high S/N peaks that dominate the constraints and in Fig.~\ref{fig:reprod_coulton} we see that the number of high peaks is suppressed by the baryonic feedback. This results in a shift toward higher values in $\Omega_\textrm{m}$, compensating for the suppression of peaks for S/N>3. We also see that the strength of the feedback is not as important as just taking it into account. 

\subsection{Further systematic effects}
\label{sec:further_effects}
At least two further effects can affect constraints: intrinsic alignment and cluster member dilution. As we briefly explain below, both effects can be suppressed, in first approximation, following \citet{Harnois_2021}, with a cut in S/N.

\subsubsection{Intrinsic alignment}

Due to the radial alignment of satellite galaxies within dark-matter halos, galaxies are not randomly oriented.\ Thus, their shape and alignments are affected by their environment and tidal fields. The intrinsic alignment has two components: the intrinsic-intrinsic correlations caused by the alignment of galaxies that are physically linked together and the gravitational-intrinsic correlations, which is the alignment of halo galaxies.  

\subsubsection{Cluster member dilution}

The source density is not homogeneous and increases around foreground clusters. Around a cluster at a given redshift, there are more galaxies but, as we do not know their redshift, they are included in the signal but may not lensed.  This effect leads to a coupling between the peak positions and the amplitude of the measured shear relative to the expected shear \citep{Kacprzak2016}. Moreover, these regions of clusters have a larger blending rate,  and thus galaxies behind clusters are more likely to be missed. These effects can result in a miscalibration between data and simulations and lead to a reduction in the mean shear signal (as we count more galaxies for the signal). An analysis of cluster member dilution in Subaru HSC weak-lensing mass maps is available, for example, in  \citet{Oguri_2021}.

\subsubsection{Reducing intrinsic alignment and cluster member dilution with a cut on S/N}

\citet{Harnois_2021} find that the effect of intrinsic alignment is small for peaks with S/N $< 3$ and that the effect of the cluster member dilution is small for S/N $< 4$. The local calibration might capture part of the cluster dilution effect, but as we have no way of knowing it without simulations, we chose to cut the S/N range to be conservative. Thus, as a first approximation of how these effects impact the cosmological parameters, we computed the constraints using only the range for peak counts of S/N $< 3$. This selection should minimize the impact of both systematic contaminations.  We tested both cases ($-2 <$ S/N $< 3$ and $-2 <$ S/N $< 6$) on mock peaks from \texttt{MassiveNuS} and find consistent results. 

\begin{figure}
    \centering
    \includegraphics[width=\hsize]{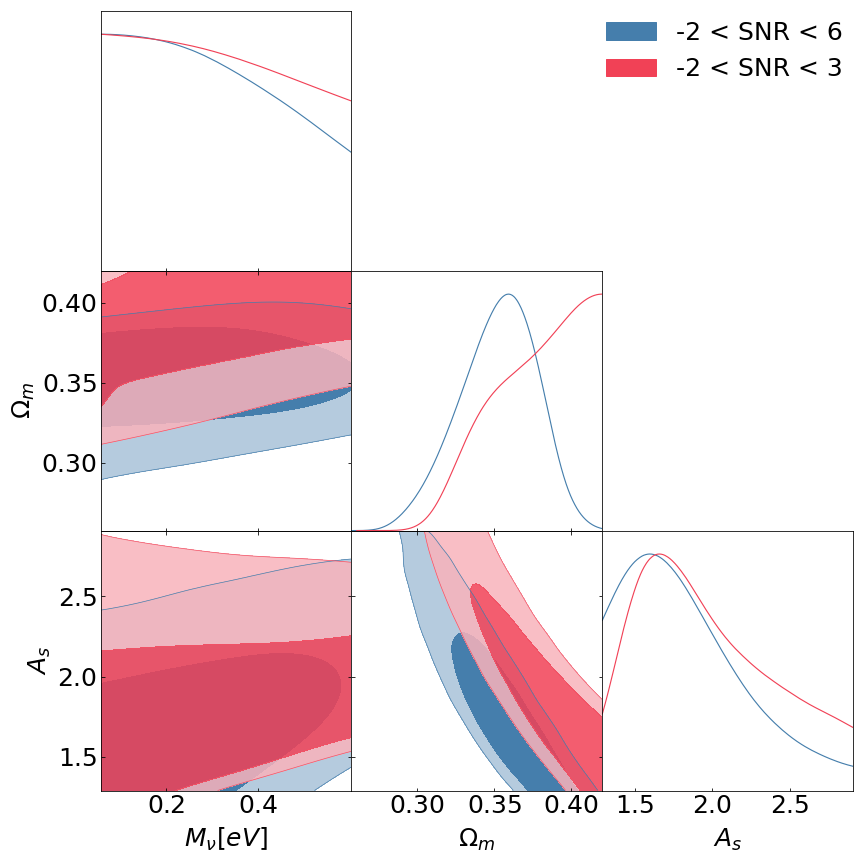}
    \caption{1D and 2D marginal posteriors for CFIS-P3 using different ranges in S/N. The 2D inner and outer contours show the $68\%$ and $95.5\%$ credible region, respectively. The shear calibration is global. The case with the full S/N range (blue) is compared to the case where we use $-2 <$ S/N $< 3$ to minimize the effect of residual bias (red). $\Omega_\textrm{m}$ shifts by $+0.027,$ which corresponds to $+0.5 \sigma$  when we cut the high S/N peaks.}
    \label{fig:constrain_end_cut}
\end{figure}

The comparison of the resulting constraints obtained with S/N $ < 3$ (red) or S/N $< 6$ (blue) is shown in Fig.~\ref{fig:constrain_end_cut}. Constraints are slightly larger when we cut the S/N range, which is expected. For both $M_\nu$ and $A_\textrm{s}$ parameters, the change is small. 

The high-S/N peaks are affected by intrinsic alignment and cluster member dilution, which leads to a reduction in peak S/N \citep{Harnois_2021}. When using the full S/N range and not accounting for those two effects, the reduction in peak S/N results in a lower inferred clustering amplitude and a lower $\Omega_\textrm{m}$. This can be seen in the $\Omega_\textrm{m}$ - $A_\textrm{s}$ 2D posterior distribution. We also perform a test using simulations only, with the range for peak counts of S/N $< 3$. As in previous tests, the data vector is the mean of the fiducial simulation at $z = 0.65$ and the result is presented in Appendix \ref{app:sim}.

\subsection{Parameter constraints that combine all systematic effects}
\label{best_case}

Our conservative model, which is the most suitable to represent the data, combines different mitigation schemes, and uses the following parameters and settings: The mean redshift is set to $\bar z = 0.65$. We calibrate the shear locally at a scale of $1$~deg$^2$. This accounts for spatial variations of shear bias and is the smallest scale for which the estimate of the selection is not dominated by noise. The residual shear bias is set to the value estimated from image simulations, $\Delta m = 0.007$. 
Baryonic feedback is accounted for by using the ``fiducial baryon'' case of \citet{Coulton_2020}. The data vector is composed of peak counts with $-2 <$ S/N $< 3$ to minimize the systematic errors from intrinsic alignment and cluster member dilution. 

\begin{figure}[ht!]
    \centering
    \includegraphics[width=\hsize]{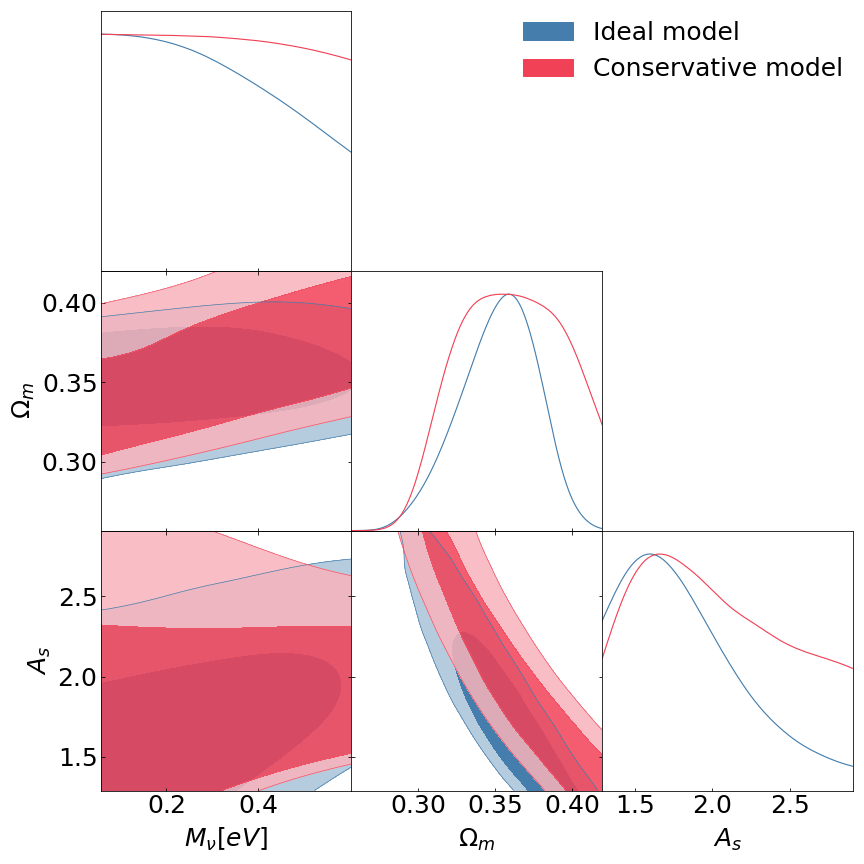}
    \caption{1D and 2D marginal posteriors for CFIS-P3 using the ideal or conservative model. The 2D inner and outer contours show the $68\%$ and $95.5\%$ credible region, respectively. The ideal one (blue) is obtained with global calibration, no baryonic correction, no intrinsic alignment, no boost factor, and on $-2 <$ S/N $< 6$. The conservative model (red) is obtained with local calibration on $1$~deg$^2$, with residual bias, modeled under the fiducial baryonic correction, and on $-2 <$ S/N $< 3$. Both models are computed at $\bar z = 0.65$. The 2D contours show the $95.5\%,$ and the 1D filled area corresponds to the constraints within the 1 sigma confidence level. $\Omega_\textrm{m}$ shifts by $+0.008,$ which corresponds to $+0.2 \sigma$ between the reference (blue) and the fiducial model (red).}
    \label{fig:all}
\end{figure}

\begin{figure*}[ht!]
    \centering
    \includegraphics[width=\textwidth]{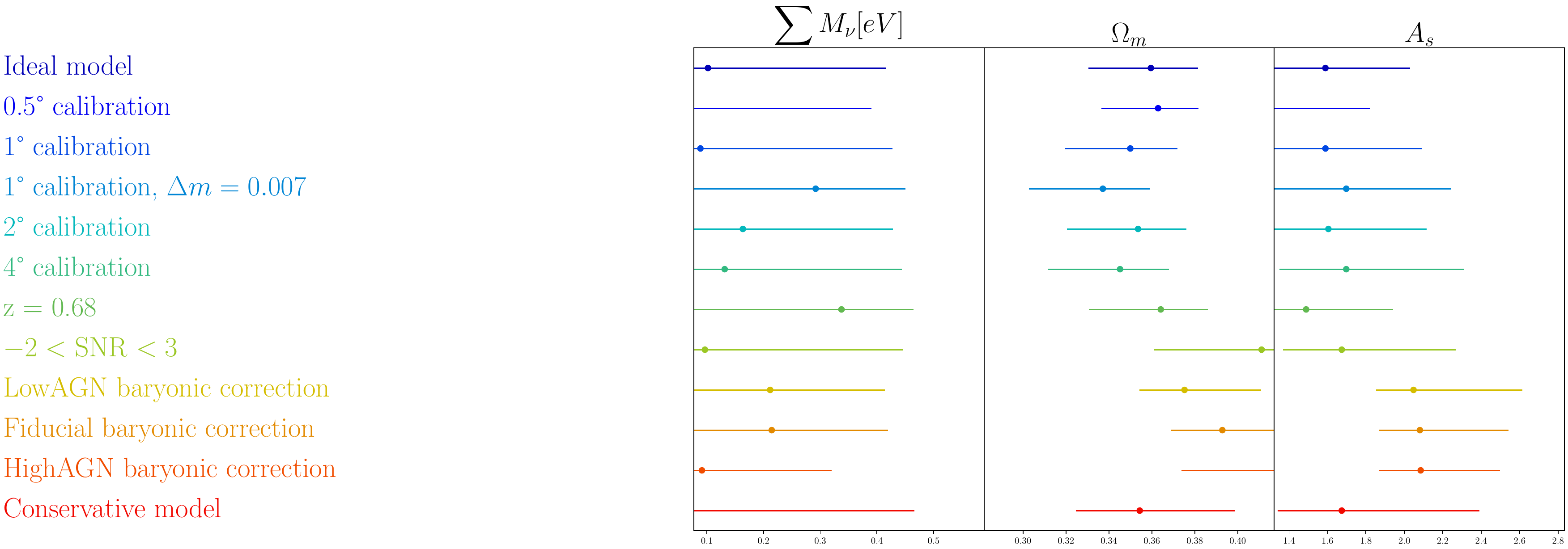}
    \caption{Marginalized distributions for the 68\% confidence interval for the three cosmological parameters for the different cases that include uncertainties and systematic effects.}
    \label{fig:summary}
\end{figure*}

In Fig.~\ref{fig:all} we show constraints using this conservative model, and compare those to the ones obtained under the ideal (optimistic) case described at the beginning of this section. $\Omega_\textrm{m}$ shifts by $+0.008,$ which corresponds to $+0.2 \sigma$ when we go from the ideal to the conservative model. As expected, the conservative model results in wider constraints. In Fig.~\ref{fig:summary}, we can see the marginalized distributions for the 68\% confidence interval of the different parameters depending on the different cases we have tested.

\section{Conclusions}
\label{conclu}

This study highlights the importance of properly accounting for systematics, such as the local calibration, residual multiplicative shear bias, intrinsic alignment and cluster member dilution, redshift uncertainties, and baryonic corrections, in the context of a peak-count cosmological analysis with CFIS data. We performed a likelihood analysis on the sum of neutrino masses, $\sum M_\nu$, the matter density parameter, $\Omega_{\rm m}$, and the amplitude of the primordial power spectrum, $A_{\rm s}$. For this purpose, we used the CFIS-P3 catalog and the \texttt{MassiveNuS} N-body simulations because they have the advantage of providing us with a large set of cosmological models and different tomographic bins. First, we obtained constraints with simulations only to validate the methodology used to model the redshift uncertainty, baryonic feedback, intrinsic alignment, and cluster member dilution. We then used the CFIS-P3 data to quantify the effect of systematics on cosmological constraints. We remark that $\Omega_\textrm{m}$ is the parameter that is constrained the most and is the most impacted by the systematics. As a summary plot, Fig.~\ref{fig:summary}, which shows the marginalized distributions for the 68\% confidence interval of the parameters per studied case. On the one hand, the baryonic corrections, the intrinsic alignment, and the cluster member dilution shift this parameter to higher values. On the other hand, a local calibration and a multiplicative shear bias $\Delta m = 0.007$ shift $\Omega_\textrm{m}$ to smaller values. Concerning $A_\textrm{s}$, the baryonic correction and the addition of $\Delta m$ shifts this parameter to higher values, whereas cutting the high S/N peaks shifts $A_\textrm{s}$ to a lower value. For $\sum M_\nu$, the parameter is not constrained enough to allow for any conclusions to be drawn. More specifically, concerning the shear calibration, it is probable that a local one is preferred over a global one, as long as the error and standard deviation are small, because it accounts for the local effects of the catalog. Nevertheless, some work has to be done to determine which calibration size is the better one. We notice that using a calibration on $1$~deg$^2$ with a value of the residual multiplicative shear bias $\Delta m = 0.007$ shifts $\Omega_\textrm{m}$ by $-0.024$ ($-0.5 \sigma$) compared to the case in which a global calibration is applied. Choosing a reasonable calibration size and having a robust estimate of $\Delta m$ is important for having a better estimate of the constraints. We note that the residual bias used here computed from image simulations neglected effects such as galaxy image blending. A more realistic estimate of $\Delta m$ is necessary for future analyses of CFIS weak-lensing data. 

Concerning the uncertainty on the mean redshift estimate, for a bias of $\Delta \bar z = 0.03,$ as considered in this study, the impact on the constraints is small enough to not be considered: the shift in $\Omega_\textrm{m}$ is only $+0.001$ ($+0.02 \sigma$). Nevertheless, we are aware that further work is needed to have a more complete description of the redshift distribution and obtain more accurate constraints. Using simulations only, the shift in $\Omega
_m$ is larger, $0.7\sigma$. This indicates that other effects in the data might lead to an underestimation of the actual shift.

To account for baryonic corrections, we considered three different flavors of baryonic feedback, labeled in this study as HighAGN, Fiducial, and LowAGN correction. We conclude that the specific flavor of baryonic feedback is less important than the difference between baryonic feedback and pure dark-matter models. Applying the LowAGN baryonic correction shifts $\Omega_\textrm{m}$ by $+0.027$ ($+0.5 \sigma$), showing the importance of modeling the baryonic feedback. Using a prediction of peak counts based on hydrodynamical simulations to model the baryonic feedback can help in getting more accurate constraints. 

Then, we see that when we minimize the effect of the cluster member dilution and intrinsic alignment by only considering peaks with S/N $< 3$, the posterior on $\Omega_\textrm{m}$ is subject to an offset of $0.027$ ($0.5 \sigma$) toward higher values. Using simulations that model these effects will significantly improve the constraints since a cut in the S/N range will no longer be necessary.

Finally, we computed a conservative model where we used the parameters and settings that are most representative of the the data. This gives larger constraints on $M_\nu$, $\Omega_\textrm{m}$, and $A_\textrm{s}$. 

We noticed that the value of $\Omega_\textrm{m}$ in particular can shift a lot due to different systematics. Our aim in this paper is not to estimate the final reference cosmological parameters, but rather to investigate, for the first time with UNIONS and peak counts,  the impact of some of the different systematics at play for this survey. Among all the systematics considered in this paper, we have shown that the one with the highest impact on $\Omega_\textrm{m}$ is related to how baryonic corrections are implemented (Fig.~\ref{fig:P3_corr_z065}); the choice of cut on the S/N (Fig.~\ref{fig:constrain_end_cut}) also causes a substantial shift. Other systematics not considered here may further shift the final parameters. It is necessary to further pursue this effort to include other systematic effects, as well as to extend the way we incorporate the baryonic feedback in the analysis and the choice of the calibration size. The constraints will be more accurate when simulations are able to also model intrinsic alignment and cluster member dilution; hydrodynamical simulations that include baryonic feedback could further improve the robustness of the results. In the future, having access to simulations built for UNIONS will allow us to make more precise investigations of various sources of bias. We also expect constraints to become more reliable with future (larger) data catalogs, such as the full UNIONS data set, for which the current pipeline will provide a starting point. 

\begin{acknowledgements}

This work is based on data obtained as part of the Canada-France Imaging Survey, a CFHT large program of the National Research Council of Canada and the French Centre National de la Recherche Scientifique.
Based on observations obtained with MegaPrime/MegaCam, a joint project of CFHT and CEA Saclay, at the Canada-France-Hawaii Telescope (CFHT) which is operated by the National Research Council (NRC) of Canada, the Institut National des Science de l’Univers (INSU) of the Centre National de la Recherche Scientifique (CNRS) of France, and the University of Hawai'i. Pan-STARRS is a project of the Institute for Astronomy of the University of Hawai'i, and is supported by the NASA SSO Near Earth Observation Program under grants 80NSSC18K0971, NNX14AM74G, NNX12AR65G, NNX13AQ47G, NNX08AR22G, and by the State of Hawai'i. 

This work has made use of the CANDIDE Cluster at the Institut d'Astrophysique de Paris and was made possible by grants from the PNCG, CNES, and the DIM-ACAV. 

This work was supported in part by the Canadian Advanced Network for Astronomical Research (CANFAR) and Compute Canada facilities.

We would like to thank the anonymous referee for their helpful suggestions that improved the manuscript.

We would also like to thank Joachim Hernois-D\'eraps for various discussions, particularly about the local metacalibration method.
We are further grateful to Jia Liu and William R.~ Coulton for useful discussions and for providing us with the baryonic corrections employed in the analysis. We would like to thank Santiago Casas for useful discussion.
Finally, we would thank the Cosmostat members for discussions about different aspects of this project. We thank the Columbia Lensing group (\url{http://columbialensing.org}) for making their simulations available. The creation of these simulations is supported through grants NSF AST-1210877, NSF AST-140041, and NASA ATP-80NSSC18K1093. We thank New Mexico State University (USA) and Instituto de Astrofisica de Andalucia CSIC (Spain) for hosting the Skies \& Universes site for cosmological simulation products.

Finally, this work uses the following software packages: \texttt{astropy, numpy, jupyter, matplotlib, sklearn, scipy, emcee, chainconsumer, getdist, os} and \texttt{sys}.

\end{acknowledgements}

\bibliographystyle{aa} 
\bibliography{bibli} 

\begin{appendix}

\section{Variation of the response matrix and the additive bias}

\begin{figure}[h!]
    \centering
    \includegraphics[width=\hsize]{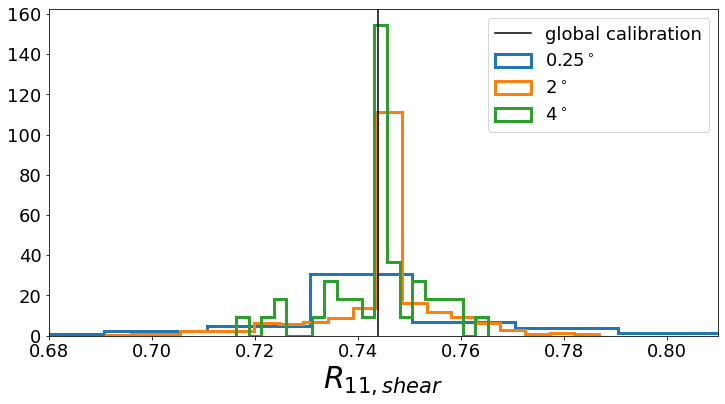}
    \includegraphics[width=\hsize]{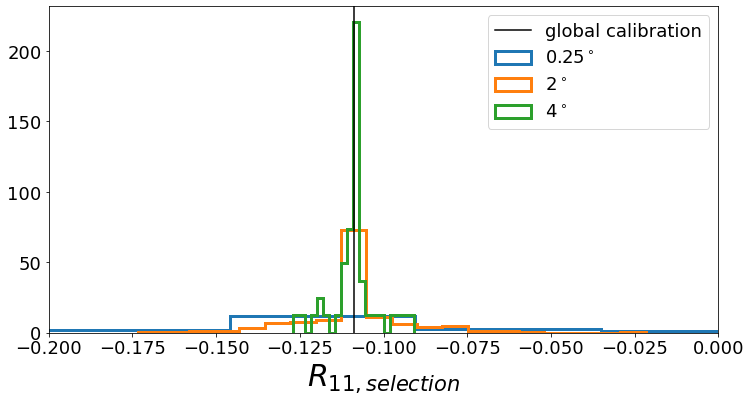}
    \caption{Histogram of $R_{11,\text{shear}}$ (top) and $R_{11,\text{selection}}$ (bottom) depending on the size of the local calibration.}
    \label{fig:hist_R}
\end{figure}

\begin{figure}[h!]
    \centering
    \includegraphics[width=\hsize]{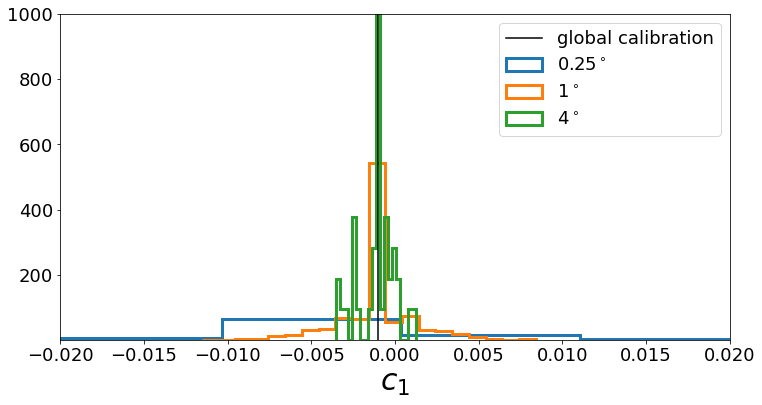}
    \includegraphics[width=\hsize]{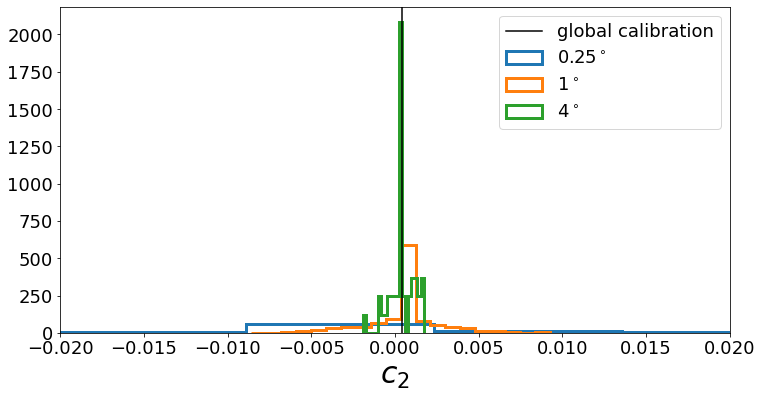}
   \caption{Histogram of $c_1$ (top) and $c_2$ (bottom) depending on the size of the local calibration.}
    \label{fig:hist_c}
\end{figure}

To have a better visualization of the variation of the response matrix and the additive bias, we compute the histogram of their distribution in Figs.~\ref{fig:hist_R} and \ref{fig:hist_c}, respectively. In every histogram, we add as a black line the value found by the global calibration. From those histograms, we can clearly see that when the pixels are smaller the dispersion is larger, which is because the number of galaxies in one patch is lower and therefore there is more noise. In all cases, we see that the values are spread around the mean one.

We confirm that the values are spread around the mean one, with larger dispersion at smaller scale of calibration.

\FloatBarrier
\section{Correlation matrix}
\begin{figure*}
    \centering
    \includegraphics[width=\textwidth]{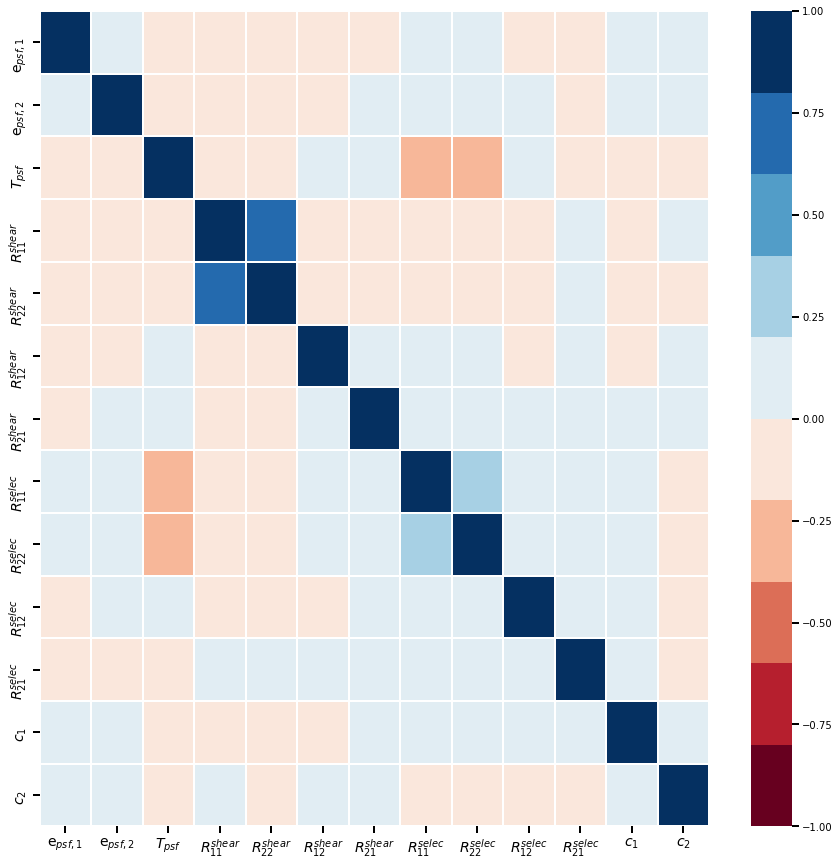}
    \caption{Correlation matrix of the quantities described in Sect.~\ref{sec:metacal} or listed in Table \ref{tab:corr}. The colors indicate the amplitude of the correlation, ranging between $-1$ and $1$.}
    \label{fig:corr_app}
\end{figure*}

The correlation matrix presented in Fig.~\ref{fig:corr_app} is presented to see the correlation between the individual elements of the matrices.  

We can observe a strong correlation (~60 - 80\%) between $R_{11}^\text{shear}$ and $R_{22}^\text{shear}$, which is expected because if the shear comports a bias, both component will be impacted.
A correlation of ~20 - 40\% between $R_{11}^\text{selection}$ and $R_{22}^\text{selection}$ is seen for the same reason as the correlation between $R_{11}^\text{shear}$ and $R_{22}^\text{shear}$. The $R_{11}^\text{selection}$ and $R_{22}^\text{selection}$ are anticorrelated (~20 - 40\%) with the size of the psf $T_\text{psf}$ because if the psf is larger, we will miss more objects. Thus, the correlation effect is stronger.
 
\newpage
\section{Impact of the systematics using only simulations} \label{app:sim}
As the data contain different systematics that are sometimes hard to model precisely, we perform some tests, using simulations only. The data vector is the mean of peak counts at the fiducial cosmology, which we modified to include systematics. The model used in the MCMC is the simulations at $z = 0.65$ without any modification.
For every test, we compare the result to a data vector without any modification (blue of every figure). 

\subsection{Redshift uncertainty}
\begin{figure}[h!]
    \centering
    \includegraphics[width=\hsize]{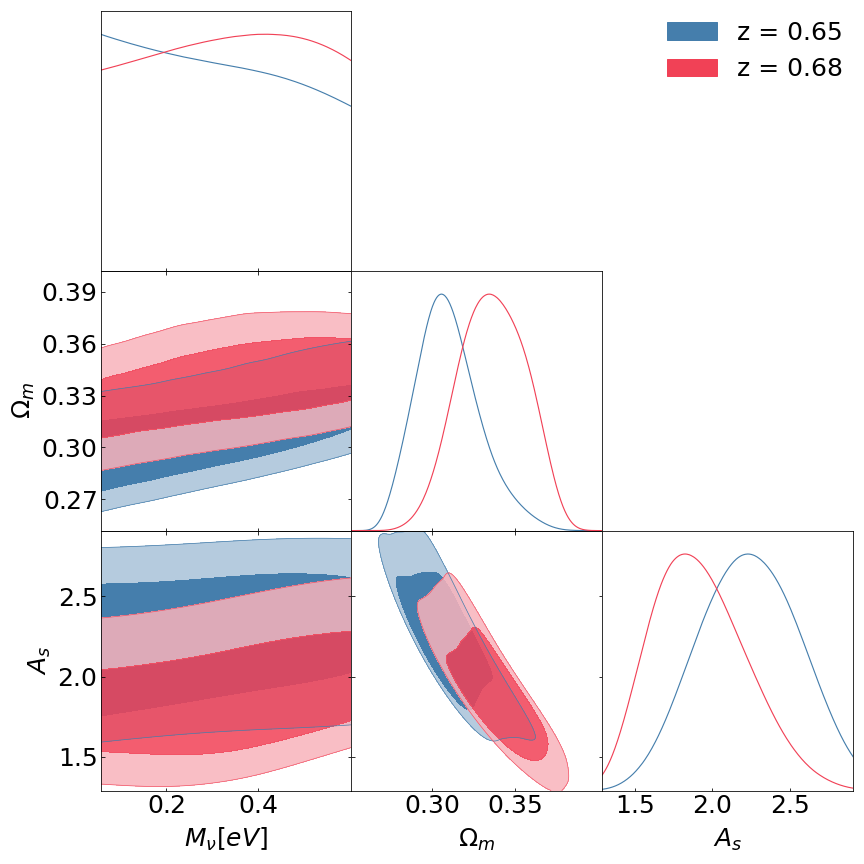}
    \caption{1D and 2D marginal posteriors using as data the mean fiducial from the simulations at different redshifts. The 2D inner and outer contours show the $68\%$ and $95.5\%$ credible region, respectively. $\Omega_\textrm{m}$ shifts by $+0.026,$ which corresponds to $+0.7\sigma$ when the redshift goes from 0.65 (blue) to 0.68 (red).}
    \label{fig:forecast_z}
\end{figure}

The impact of the redshift uncertainty is shown in Fig.~\ref{fig:forecast_z}, where we compute the data vector at $z = 0.65$ (blue) or $z = 0.68$ (red), whereas the model is kept at $z = 0.65$ in both cases. Modeling with lower redshift shifts $\Omega_m$ by $+0.7 \sigma$.

\subsection{Intrinsic alignment and cluster member dilution}
\begin{figure}[h!]
    \centering
    \includegraphics[width=\hsize]{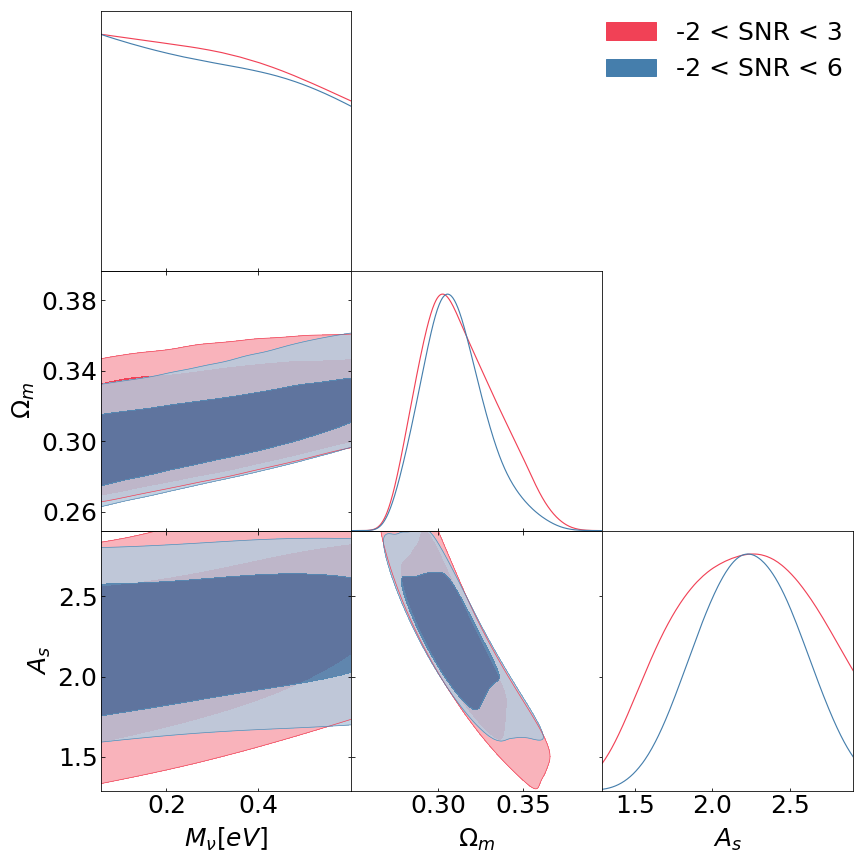}
    \caption{1D and 2D marginal posteriors using as data the mean fiducial from the simulations and different ranges in S/N. The 2D inner and outer contours show the $68\%$ and $95.5\%$ credible region, respectively. The shear calibration is global. The case with the full S/N range (blue) is compared to the case where we use $-2 <$ S/N $< 3$ to minimize the effect of residual bias (red). The constraints are larger when we cut the high S/N peaks.}
    \label{fig:forecast_cut}
\end{figure}

The method used here to decrease the impact of intrinsic alignment and cluster member dilution is to cut the high S/N range. In Fig.~\ref{fig:forecast_cut}, we show the constraints when we use a S/N range between $-2$ and $6$ (blue) or a reduced S/N range between $-2$ and $3$ (red). When the reduced range is used, the constraints are larger, which is expected because there is less information.

\end{appendix}
\end{document}